\documentclass[prb, preprint, twocolumn, amsmath, amssymb, reprint, superscriptaddress]{revtex4-1}

\usepackage{graphicx,verbatim}
\usepackage{dcolumn}
\usepackage{bm}
\usepackage{sidecap}
\usepackage{braket}
\usepackage{color}

\begin{document}

\title{Current-induced bond rupture in single-molecule junctions}

\author{A. Erpenbeck}
\affiliation{
Institut f\"ur Theoretische Physik und Interdisziplin\"ares Zentrum f\"ur Molekulare 
Materialien, \\
Friedrich-Alexander-Universit\"at Erlangen-N\"urnberg,\\ 
Staudtstr.\, 7/B2, D-91058 Erlangen, Germany
}
\author{C. Schinabeck}
\affiliation{
Institut f\"ur Theoretische Physik und Interdisziplin\"ares Zentrum f\"ur Molekulare 
Materialien, \\
Friedrich-Alexander-Universit\"at Erlangen-N\"urnberg,\\ 
Staudtstr.\, 7/B2, D-91058 Erlangen, Germany
}
\affiliation{
Physikalisches Institut, Albert-Ludwigs-Universit\"at Freiburg, \\
Hermann-Herder-Strasse 3, 79104 Freiburg, Germany
}
\author{U.\ Peskin}
\affiliation{
Schulich Faculty of Chemistry, Technion-Israel Institute of
Technology, Haifa 32000, Israel
}
\author{M.\ Thoss}
\affiliation{
Institut f\"ur Theoretische Physik und Interdisziplin\"ares Zentrum f\"ur Molekulare 
Materialien, \\
Friedrich-Alexander-Universit\"at Erlangen-N\"urnberg,\\ 
Staudtstr.\, 7/B2, D-91058 Erlangen, Germany
}
\affiliation{
Physikalisches Institut, Albert-Ludwigs-Universit\"at Freiburg, \\
Hermann-Herder-Strasse 3, 79104 Freiburg, Germany
}

\date{\today}

\begin{abstract}
		Electronic-vibrational coupling in single-molecule
		junctions may result in current-induced bond rupture and is 
		thus an important mechanism for the stability of molecular junctions. We use the hierarchical quantum master equation (HQME) method in combination with the quasi-classical Ehrenfest 
		approach for the nuclear degrees of freedom to simulate current-induced bond rupture in single-molecule junctions. Employing 
		generic models for molecular junctions with dissociative nuclear
		potentials, we analyze the underlying mechanisms. 
		In particular, we investigate the dependence of the 
		dissociation probability on the applied bias voltage and the molecule-lead coupling strength.
		The results show that an applied bias voltage can not only lead to dissociation of the molecular junction, but under certain conditions can also increase the stability of the molecule.
\end{abstract}


\maketitle

\section{Introduction}

	Nonequilibrium quantum physics in nanostructures is an active field of research. Among the systems investigated are molecular junctions, which comprise a single molecule attached to two macroscopic leads at finite bias voltage. They provide a versatile architecture to study fundamental aspects of nonequilibrium quantum physics at the nanoscale and are of interest for applications in the field of molecular scale electronics.\cite{Nitzan2001, Nitzan2003, Cuniberti, Galperin_Vib_Effects, Cuevas_Scheer, Zimbovskaya2011, Bergfield2013, Baldea, Su2016,Thoss2018}

	The coupling between the current-induced charge-fluctuations and the nuclear (vibrational) degrees of freedom plays an essential role in molecular junctions.\cite{Galperin_Vib_Effects,Rainer_Baldea, Galperin2005, Leijnse2008, Rainer2012, Rainer2011, Rainer2013, Wilner2014, Schinabeck2014, Erpenbeck2015}
	Experimental as well as theoretical studies have shown that a current across a molecular junction induces nonequilibrium vibrational excitation.\cite{Ioffe2008, Schulze2008, deLeon2008, Huettel2009, Rainer2009, Rainer2010, Ward2010, Rainer2011, Rainer2011b, Schinabeck2016} While the level of current-induced vibrational excitation is typically small for low voltages, which corresponds to the off-resonant transport regime, it can be substantial for higher voltages, in particular in the resonant transport regime. In that regime, current-induced heating can cause mechanical instability of the junction and may eventually result in bond rupture, i.e. dissociation of the molecule.	
	This process of current induced bond rupture  has recently been observed  experimentally in molecular junctions.\cite{Venkataraman2015, Sabater2015, Venkataraman2016, Capozzi2016} 
	The fact that stable molecular junctions are rarely observed for voltages larger than $\sim 1-2$ V is a further indication for the relevance of this process. The understanding of the underlying mechanisms of bond rupture and its implication for the stability in molecular junctions is thus not only of fundamental interest in the fields of nonequilibrium nanophysics, but is also crucial for the design of molecular junctions, which are stable at higher voltages.
	
	It is noted that similar processes have also been investigated in the field of surface science. For example, studies using scanning tunneling microscope (STM) setups, where a cantilever injects electrons into a molecule on a surface, have revealed that a current through a molecule can lead to desorption from the surface.
	\cite{Gao1995_1, Gao1995_2, Avouris1996, Gao1997, Boendgen1998, Seideman2003, Saalfrank2006, Menzel2012}
	Moreover, STM experiments found that an electric current can break\cite{Martel1996, Stipe1997, Lauho2000, Hla2003, Roy2013} or form\cite{Lee1999} molecular bonds at surfaces.
	Depending on the details of the setup and the molecules under investigation, there are several different processes that can cause these effects, such as current-induced vibrational excitation\cite{Stipe1997, Seideman2003} or the population of an excited, possibly anti-bonding, electronic state.\cite{Avouris1996, Boendgen1998, Seideman2003}
	Similar processes were also considered in molecular dissociation and desorption from a surface upon laser excitation.\cite{Gao1995_1, Gao1995_2, Brandbyge1995, Saalfrank1996, Vondrak1999}
	Theoretical approaches to study these mechanisms at surfaces range from the description of the nuclear reaction coordinate in terms of truncated harmonic oscillators\cite{Gao1995_1, Brandbyge1995, Stipe1997, Saalfrank2006}
	and Morse potentials\cite{Brandbyge1995, Koch2006, Roy2013} 
	to quasi-classical wave-packet dynamics\cite{Avouris1996, Boendgen1998} 
	and quantum mechanical approaches using spatial grid representations.\cite{Saalfrank1996, Boendgen1998}

	In the context of molecular junctions, the theoretical framework to study current-induced vibrational excitation is well established for models, which treat the vibrational modes within the harmonic approximation.\cite{Wegewijs2005, Ryndyk2006, Benesch2008, Rainer2009, Rainer2011, Erpenbeck2016, Schinabeck2016} 
	While such models have been used to investigate the mechanical stability of molecular junctions,\cite{Koch2006, Rainer2010, Volkovich2011, Rainer2011, Rainer2015} the study of bond rupture requires to go beyond the harmonic approximation and use nuclear potentials which can describe the dissociation process explicitly. So far, this has been achieved within a classical treatment of the nuclei \cite{Dzhioev2011, Dzhioev2013, Pozner2014} or using perturbative rate theories.\cite{Koch2006, Brisker2008}
	
	In this paper, we study voltage-induced bond rupture in single-molecule junctions based on generic model systems using a mixed quantum-classical approach to transport. In this approach, the electrons are treated fully quantum-mechanically within the numerically exact hierarchical quantum master equation (HQME) approach.\cite{Tanimura1989, Tanimura2006, Welack2006, Popescu2013, Haertle2013a, Haertle2014, Haertle2015, Wenderoth2016, Jin2008, Li2012, Zheng2013, Cheng2015, Ye2016, Schinabeck2016, Cheng2017, Hou2017} The nuclear motion, on the other hand, is described by the classical Ehrenfest method.\cite{Stock2005, Horsfield2004, Verdozzi2006, Todorov2010, Subotnik2010, Todorov2014, Kartsev2014, Cunningham2015, Bellonzi2016}
	The use of the HQME method allows to solve the transport problem for a dissociative system within the Ehrenfest approximation without further approximation, thus extending previous related methodologies.\cite{Hussein2010, Metelmann2011,Dzhioev2013}
	Applying this approach, we study the effect of voltage-induced bond rupture for a wide range of model parameters, ranging from the nonadiabatic regime of weak molecule-lead coupling to the adiabatic case of strong coupling.
	
	The outline of the paper is as follows: In Sec.\ \ref{sec:theory} we introduce the model and the theoretical approach. In Sec.\ \ref{sec:results} we show results for representative model systems and give a systematic overview of effects associated with voltage-induced bond rupture. Thereby, we distinguish three different scenarios for the coupling between the molecule and the leads. Sec.\ \ref{sec:conclusion} concludes with a summary.

\section{Theoretical methodology} \label{sec:theory}

\subsection{Model}

	In order to investigate current-induced bond rupture in single-molecule junctions, we consider a model system consisting of a molecule coupled to two macroscopic leads described by the Hamiltonian
	\begin{eqnarray}
		H	&=&	H_{\text{M}} + H_{\text{ML}} + H_{\text{MR}} + H_{\text{L}} + H_{\text{R}}.
	\end{eqnarray}
	The Hamiltonian of the molecule is given by
	\begin{eqnarray}
		H_{\text{M}} 	&=&	\frac{p^2}{2m} + V_0(x) (1-d^\dagger d) + V_d(x) d^\dagger d.
	\end{eqnarray}
	It describes a single electronic state, which can be empty (in the following referred to as the neutral state of the molecule) or occupied (charged state), coupled to a nuclear degree of freedom $x$ along which the molecule can dissociate.  Thereby, $d^{\dagger}$/$d$ denote the electronic creation/annihilation operators, respectively; $p$ is the momentum and $m$ the reduced mass of the nuclear mode. 
	Within this model, $V_0(x)$ and $V_d(x)$ describe the nuclear potential energy surfaces of the neutral and the charged state of the molecule, respectively. In the following, we will assume that $V_0(x)$ is a bonding and $V_d(x)$ is an anti-bonding potential. The specific potentials used will be specified in Sec.\ \ref{sec:results}. In this paper, we use a description in reduced dimensionality, focusing on a single nuclear degree of freedom describing the dissociation of a molecular bond.

	To allow for electron transport, the molecule couples to two macroscopic leads which are modeled as reservoirs of non-interacting electrons,
	\begin{eqnarray}
		H_{\text{L/R}} 	&=&	\sum_{k \in {\text{L/R}}} \epsilon_k c_k^\dagger c_k.
	\end{eqnarray}
	Here, $\epsilon_k$ is the energy of lead-state $k$ and $c_k^{\dagger}$/$c_k$ the corresponding creation/annihilation operators.
	The interaction between the molecule and the leads is given by
	\begin{eqnarray}
		H_{\text{ML/R}}	&=&	\sum_{k \in {\text{L/R}}} V_{k}(x) c_k^\dagger d + \text{h.c.}\ .
	\end{eqnarray}
	The coupling parameters $V_{k}(x)$ are described by the 
	spectral density for the interaction between the molecule and the leads,
	\begin{eqnarray}
		\Gamma_{\text{L/R}}(x, \epsilon)	&=&	2\pi \sum_{k \in {\text{L/R}}} |V_{k}(x)|^2 \delta(\epsilon_k-\epsilon) . \label{eq:def_Gamma}
	\end{eqnarray}
	The coupling between the molecule and the leads determines the conduction properties of the molecular junction. The position-dependent coupling $V_{k}(x)$ allows to model the situation where the molecular conductance depends on the nuclear degree of freedom (see below). This is important if the conductance of the molecule changes upon dissociation of molecular bonds.
	In the results reported in the following, we will exclusively work in the wide-band limit, that is the spectral density is energy independent.

\subsection{Transport theory}

	We use a mixed quantum-classical approach to describe transport across the molecular junction. The electrons are treated fully quantum-mechanically within the numerically exact hierarchical quantum master equation (HQME) approach. The nuclear motion, on the other hand, is described by the classical Ehrenfest method. 
	It is noted, though, that we solve the transport problem for a dissociative system within the Ehrenfest approach without further approximations. Thus, we go beyond previous work that considered transport within the Ehrenfest approach, but considered either harmonic nuclear degrees of freedom\cite{Hussein2010, Metelmann2011} or applied a separation of timescales approximation to study dissociative systems.\cite{Dzhioev2013}
	We briefly discuss the HQME and the Ehrenfest approach in the following.
	
\subsubsection{Electron dynamics}
	The HQME approach, also known as hierarchical equation of motion (HEOM) approach, was originally developed by Tanimura and Kubo to describe relaxation dynamics in quantum systems,\cite{Tanimura1989, Tanimura2006}
	but also allows for a description of nonequilibrium electron transport in quantum systems. 
 	\cite{Welack2006, Popescu2013, Haertle2013a, Haertle2014, Haertle2015, Wenderoth2016, Jin2008, Li2012, Zheng2013, Cheng2015, Ye2016, Schinabeck2016, Cheng2017, Hou2017} 
	As a numerically exact approach, the HQME framework does not suffer from the usual limitations of perturbative approaches, that is being limited to at least one weak coupling parameter. For a detailed derivation of the HQME method in the context of quantum transport, we refer to Refs.\ \onlinecite{Yan2008, Zheng2012, Haertle2013a}.
	
	The HQME is an approach to the dynamics of open quantum system. Accordingly, the overall problem is separated into a system and a bath. In the molecular junction scenario considered here, the leads represent the bath, while the molecule, including the electronic state and the nuclear degree of freedom, constitute the system.  The HQME theory provides an equation of motion for the reduced density matrix of the system, $\rho(t)$, given by
		\begin{eqnarray}
			\frac{\partial}{\partial t} \rho(t) &=& -\frac{i}{\hbar} [H_{\text{M}} , \rho(t)]  \label{eq:EQM_0th_tier}
			\\&&- \frac{i}{\hbar^2} 
			\hspace*{-0.15cm}
			\sum_{K \in \lbrace{\text{L}, \text{R}}\rbrace\atop p \in {\text{poles}}}
			\hspace*{-0.2cm}
			V_{K}(x) \left( [d, \rho_{Kp+}^{(1)}(t)] + [d^\dagger, \rho_{Kp-}^{(1)}(t)]\right) \nonumber . 
		\end{eqnarray}
	Thereby, $\rho_{Kp\pm}^{(1)}(t)$ denote $1$st-tier auxiliary density operators. In general, there is an infinite hierarchy of $n$th-tier auxiliary density operators $\rho_{a_1 \dots a_n}^{(n)}(t)$, which obey the equation of motion
		\begin{eqnarray}
			\frac{\partial}{\partial t} \rho_{a_1 \dots a_n}^{(n)} 	&=& 
			-\frac{i}{\hbar} [H_\text{M}, \rho_{a_1 \dots a_n}^{(n)}]
			-\left( \sum_{j=1}^n \gamma_{a_j} \right) \rho_{a_1 \dots a_n}^{(n)} \nonumber \\&&
			-i \sum_{j=1}^n (-1)^{n-j} \mathcal{C}_{a_j} \rho_{a_1 \dots a_{j-1} a_{j+1} \dots a_n}^{(n-1)} \nonumber \\&&
			-\frac{i}{\hbar^2} \sum_{a_{n+1}} A_{K_{n+1}}^{\overline{\sigma_{{n+1}}}} \rho_{a_1 \dots a_n a_{n+1}}^{(n+1)}.
			\label{eq:EQM_nth_tier}
		\end{eqnarray}
	They describe the influence of the leads on the dynamics of the molecule. The $n$th-tier auxiliary density matrices have $n$ compound indices $a_j = (K_j, p_j, \sigma_j)$, consisting of a lead index $K_j\in \lbrace \text{L}, \text{R} \rbrace$ and an index corresponding to the molecular creation/annihilation operator $\sigma_j \in \lbrace +,- \rbrace$. 
	Within the HQME approach, the influence of the environment is encoded in the two-time correlation function of the free bath defined as
	\begin{eqnarray}
		C_K^\pm(t, t') &=& \sum_{k \in K} V_k(x(t')) \braket{ F^\pm_{Kk}(t) F^\mp_{Kk}(t') },
	\end{eqnarray}
	with the operators 
	\begin{eqnarray}
		F^\pm_{Kk}(t) &=& \exp\left(\frac{i}{\hbar} H_K t\right) c_k^\pm \exp\left(-\frac{i}{\hbar} H_K t \right)
	\end{eqnarray}
	and $c_k^- = c_k$ and $c_k^+ = c_k^\dagger$.
	Notice that we are using a slightly different definition of the correlation function compared to, for example, Refs.\ \onlinecite{Welack2006, Yan2008, Haertle2013a, Schinabeck2016}, as this simplifies the treatment of non-constant molecule-lead couplings.
	The index $p_j\in \mathbb{N}$ stems from the decomposition of this correlation function of the free bath in terms of exponentials which allows for a systematic closure of the equations entering the hierarchy.\cite{Yan2008, Zheng2012, Haertle2013a}
	For the wide-band limit considered in this work, the specific decomposition is given by
	\begin{subequations}
		\begin{eqnarray}
	 \hspace*{-1cm}
	 C_K^\pm(t, t')	&=&	
				\int d\epsilon\  e^{\pm \frac{i}{\hbar}\epsilon (t-t')} \ V_{K}(x(t'))\ f(\pm\epsilon, \pm\mu_K) \\
			&=& 
					  \hbar \pi \ V_{K}(x(t')) \ \delta(t-t')	\nonumber \\&&
					- \sum_{p=1}^\infty \frac{2i\pi V_{K}(x(t'))}{\beta} \ \eta_p \ e^{-\gamma_{Kp\pm} (t-t')}, \label{eq:decomposition}
		\end{eqnarray}
	\end{subequations}
	with the Fermi distribution function $f(\epsilon, \mu) = \left( 1 + \exp(\beta(\epsilon-\mu)\right)^{-1}$ with $\mu$ being the chemical potential, $\beta = \frac{1}{k_B T}$ with Boltzmann constant $k_B$ and temperature $T$. 
	The HQME (\ref{eq:EQM_0th_tier}) and (\ref{eq:EQM_nth_tier}) contain the objects
	\begin{subequations}
		\begin{eqnarray}
	\gamma_{a}				&=& - \sigma \frac{i}{\hbar} \left(\mu_K + \frac{i \sigma \chi_p}{\beta} \right)  , \\ 
	\mathcal{C}_{a}	\rho^{(n)}		&=& - \frac{2i\pi V_{K}(x)}{\beta}  \eta_p   \left\lbrace d^{\sigma}, \rho^{(n)} \right\rbrace_{(-1)^{n+1}} ,  \\
	A_{K}^{\overline{\sigma}} \rho^{(n)} 	&=& V_{K}(x) \left\lbrace d^{\overline{\sigma}}, \rho^{(n)} \right\rbrace_{(-1)^{n}} ,
		\end{eqnarray}
	\end{subequations}
	where $\overline{\sigma}= - \sigma$, $d^- = d$ and $d^+ = d^\dagger$. Thereby, $\lbrace . , . \rbrace_-$ denotes the commutator, $\lbrace . , . \rbrace_+$ is the anti-commutator. These expressions are specific for the wide-band limit and the Pade decomposition \cite{Hu2010, Hu2011} used throughout this paper. How to calculate the Pade decomposition parameters $\eta_p$ and $\chi_p$ was for example demonstrated by \citet{Hu2011}
	
	The $\delta$-function in Eq.\ (\ref{eq:decomposition}) is characteristic for the wide-band limit. It needs to be treated differently than the sum over exponentials.\cite{Croy2009,Zheng2010, Zhang2013,Kwok2014}
	In order to consistently include the $\delta$-function in the equations of motion (\ref{eq:EQM_0th_tier}) and (\ref{eq:EQM_nth_tier}), we extend the index set of poles $p$ by zero.
	The auxiliary density operators corresponding to $p=0$ are not obtained by forward propagation of the differential equations (\ref{eq:EQM_nth_tier}), instead they are calculated as
	\begin{eqnarray}
	  \rho_{a_1 \dots a_n (K, 0, \sigma)}^{(n+1)} 	&=&	- \frac{i\pi\hbar V_{K}(x)}{2} 
									 \cdot \left\lbrace d^{\sigma}, \rho_{a_1 \dots a_n}^{(n)} \right\rbrace_{(-1)^{n+1}} . \nonumber\\
	\end{eqnarray}

	As can be seen in Eq.\ (\ref{eq:EQM_nth_tier}), the equations of motion for the $n$th-tier auxiliary density operators couple to the $(n+1)$th-tier via the operator $A_{K}^{\overline{\sigma}}$ and to the $(n-1)$th-tier via $\mathcal{C}_{a}$. In general, this results in an infinite hierarchy of coupled differential equations, which has to be truncated in a suitable manner for applications.\cite{Tanimura1991, Yan2004, Xu2005, Schroeder2007} 
	As we are only interested in the molecular population and the electronic current, which are single particle observables (see below) and describe the nuclear motion classically such that the electronic system is effectively non-interacting, the hierarchy terminates after the $2$nd-tier.\cite{Yan2008, Karlstrom2013}
	Within the wide-band limit, it is sufficient to only include the $1$st-tier auxiliary density matrices and still obtain numerically exact results.\cite{Croy2009, Zheng2010, Kwok2014, Leitherer2017}

\subsubsection{Nuclear dynamics}\label{sec:nuc_dynamics}
	We describe the dynamics of the nuclear degree of freedom classically within the Ehrenfest approach.\cite{Stock2005, Horsfield2004, Verdozzi2006, Todorov2010, Subotnik2010, Todorov2014, Kartsev2014, Cunningham2015, Bellonzi2016}
	Within this approach, the electrons act on the nuclear degrees of freedom via the mean force and the equations of motion for the position $x$ and momentum $p$ of the classical trajectory read
	\begin{subequations}
		 \begin{eqnarray}
			m \dot x		&=&	p,  \label{eq:Ehrenfest_x}  \\
			\dot p 		&=& 	- \text{Tr} \left\lbrace \rho \frac{\partial H}{\partial x} \right\rbrace \nonumber\\
						&=&	- \left(  \rho_{00} \frac{\partial V_0 (x)}{\partial x} + \rho_{11} \frac{\partial V_d (x)}{\partial x}  \right)  \\
						&&	- \sum_{K \in \lbrace{\text{L}, \text{R}}\rbrace\atop p \in {\text{poles}}}
							\frac{\partial V_{K}(x)}{\partial x} 
							\text{Tr} \left\lbrace 
							\rho^{(1)}_{Kp+} d + d^\dagger \rho^{(1)}_{Kp+}
							\right\rbrace. 		\nonumber					
		\end{eqnarray}\label{eq:Ehrenfest_p}
	\end{subequations}
	Thereby, $\text{Tr} \left\lbrace . \right\rbrace$ denotes the trace over the electronic degree of freedom at the molecular bridge. $\rho_{00}$ and $\rho_{11}$ are the diagonal elements of the reduced density matrix, representing the probability that the electronic state is empty or populated, respectively.
	Within the mixed quantum-classical Ehrenfest approach, the initial quantum state of the nuclear degrees of freedom is modeled by a sampling of the initial values of the classical trajectories using an appropriate phase-space distribution, e.g., the Wigner function of the initial state.\cite{Stock2005} 
	In the calculations reported below, we have used a Gauss-Hermite quadrature \cite{Numerical_recipes} to sample the Wigner function 
		\begin{eqnarray}\label{wigner}
			 \rho_W(x,p) &=& \frac{1}{\pi\hbar} \tanh\left( \frac{\hbar\omega}{2k_BT}\right) \times \nonumber \\&&
					 \times e^{  -\tanh\left( \frac{\hbar\omega}{2k_BT}\right) \left( \frac{m \omega}{\hbar}(x-x_0)^2 + \frac{1}{m\hbar\omega}p^2 \right)},
		\end{eqnarray}
	which corresponds to the thermal equilibrium of the neutral state of the molecule.	The frequency $\omega$ is thereby determined by the harmonic approximation to the potential $V_0(x)$ at its minimum $x_0$.
	The sampling provides the initial values, $x_j(0)$ and $p_j(0)$, of the classical trajectories, $x_j(t)$ and $p_j(t)$, which are then obtained solving the equations of motion (\ref{eq:Ehrenfest_p}) using a fourth-order Runge-Kutta method. The weight of each trajectory, $P_j$, is determined by the phase-space distribution (\ref{wigner}).
	
	Describing the nuclear degree of freedom by classical trajectories is expected to be a valid approximation for a sufficiently large reduced mass $m$. 
	Even though the Ehrenfest method has been used to assess current-induced nuclear motion,\cite{Horsfield2004_3, Verdozzi2006, Dundas2009, Cunningham2015} it is also known that some physical effects can not be described by this approach, e.g.\ Joule heating.\cite{Horsfield2004_2, Horsfield2004_3} 
	However, these effects are not essential for the studies in this paper, which focus on current-induced bond rupture upon the transient population of anti-bonding molecular electronic states by tunneling electrons.

\subsubsection{Observables}
	Several observables are of interest for analysis of the transport problem. The most important observable to study bond rupture is the long-time dissociation probability given by
		\begin{eqnarray}
			P_{\text{total}} (t)	&=&	
							\hspace*{-0.25cm}
							\sum_{j\in\text{trajectories}} 
							\hspace*{-0.25cm}
							P_j \cdot \theta(x_j(t) - x_{\text{threshold}}) , \label{eq:diss_probab}
		\end{eqnarray}
		where $\theta$ is the Heaviside step function.
		Thereby, a trajectory $x_j(t)$ is counted as dissociated whenever it exceeds a certain threshold value $x_{\text{threshold}}$. In the calculations reported below, we have used a threshold valued of $x_{\text{threshold}}= 5\AA$, after test calculations. The specific value of $x_{\text{threshold}}$ only influences the short-time dynamics of $P_{\text{total}} (t)$, while the long time limit $P_{\text{total}} (t\rightarrow\infty)$ is insensitive to (reasonable) choices of $x_{\text{threshold}}$.

	Another important observable is the current. Within the HQME framework, the current between lead L/R and the molecule for trajectory $j$ is calculated as
		\begin{eqnarray}
			I_{j\text{ L/R}}	&=&	\frac{ie}{\hbar^2} \sum_{K \in {\text{L/R}}\atop p \in {\text{poles}}} V_{K}(x_j) \text{Tr}\left( d \rho_{Kp+}^{(1)} - d^\dagger \rho_{Kp-}^{(1)} \right) \label{eq:current} .
		\end{eqnarray}
	In the following, we will study the total current given as the average over all trajectories,
		\begin{eqnarray}
			I_{\text{L/R}}	&=&	\hspace*{-0.25cm}
						\sum_{j\in\text{trajectories}} 
						\hspace*{-0.25cm}
						P_j I_{j\text{ L/R}}.
		\end{eqnarray}
	This current corresponds to an average over many repetitions of an experiment, where both stable and dissociated molecular junctions contribute. Notice that an individual molecular junction can either be stable or dissociated.

\section{Results} \label{sec:results}

	In the following, we apply the methodology introduced above to analyze bond rupture induced by an applied bias voltage. After an outline of the model parameters and some details on the simulations in Sec.\ \ref{sec:specification}, we consider three different molecule-lead coupling scenarios in Secs.\ \ref{sec:gate-voltage} --- \ref{sec:asymmetric}. To stay within the range of validity of the Ehrenfest approach, we thereby focus on the resonant transport regime, where molecular dissociation is a consequence of the transient population of anti-bonding states by tunneling electrons rather than by heating of vibrational modes.

	\subsection{Specification of model parameters and details of simulation}\label{sec:specification}

\begin{figure}[htb!]
		\includegraphics[width = 0.33\textwidth]{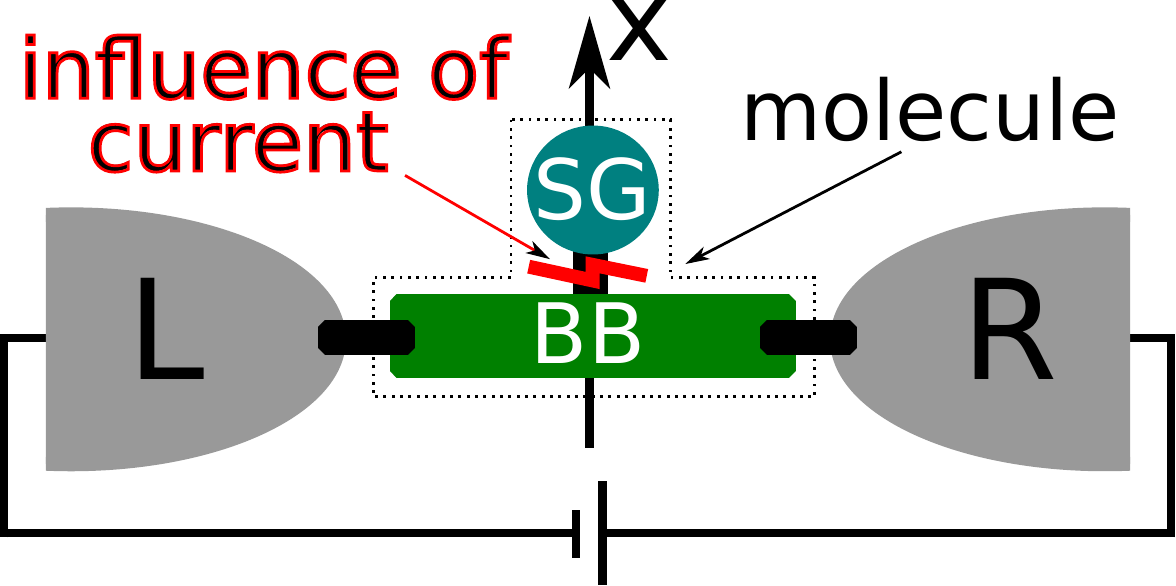}
		\caption{Sketch of the model under investigation, which exhibits current-induced non-destructive dissociation. The molecular junction consists of a backbone (BB) and a side-group (SG).}
		\label{fig:sketch}
	\end{figure}

	The model and transport formalisms introduced above are applicable to different scenarios of current-induced bond rupture in molecular junctions. In this work, we focus on non-destructive current-induced bond rupture in single-molecule junctions. To this end, we specifically consider a scenario schematically depicted in Fig.\ \ref{fig:sketch}. The molecular bridge consists of a backbone (BB) and a side-group (SG). We model the system in such a way, that the current through the molecule influences the bond between the side-group and the backbone. If the current leads to bond rupture, the side-group will detach from the backbone and dissociate ($x\rightarrow\infty$). In this scenario, the leads remain bridged by the molecular backbone, thus we refer to this mechanism as non-destructive. A similar model has already been used to investigate bond dissociation induced by charge fluctuation in a donor-bridge-acceptor complex.\cite{Brisker2008}

\begin{figure}[htb!]
			\includegraphics[width = \linewidth]{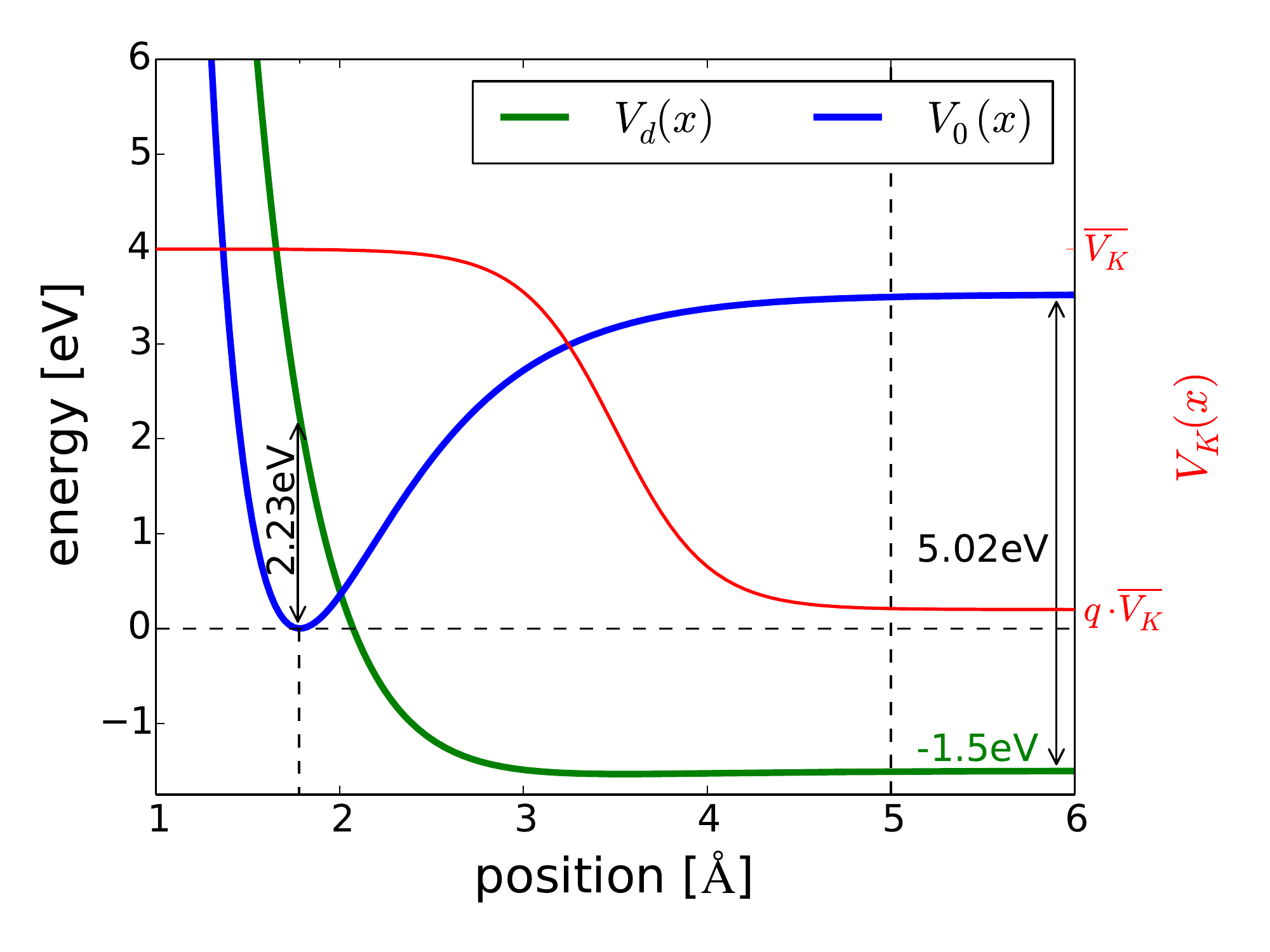}
			\caption{Potential energies used to describe the bond between the backbone and the side-group of the molecular bridge in the neutral ($V_0(x)$) and the charged state ($V_d(x)$). The red line visualizes the dependence of the molecule-lead coupling on the nuclear coordinate.}
			\label{fig:pots_overview}
		\end{figure}

	The nuclear potentials of the neutral and the charged state are assumed to be bonding and anti-bonding, respectively, along the dissociation coordinate and are depicted in Fig.\ \ref{fig:pots_overview}.
	Specifically, for the neutral molecule, the bond between the backbone and the side-group along the $x-$axis is described by the binding Morse-potential 
	\begin{eqnarray}
		V_0(x) &=& 	D_e \cdot\left( e^{-a(x-x_0)} -1 \right)^2 + c,
	\end{eqnarray}
	where $x_0=1.78\AA$ is the equilibrium bond distance, $D_e = 3.52\text{eV}$ the dissociation energy, $a=1.7361/\AA$ the width of the Morse potential, resulting in $\hbar\omega = 91.7$meV, and $c= -45.7\text{meV}$ a constant shifting the absolute energy of the potential. 
	The parameter $c$ was for convenience chosen such that the energy of the quantum mechanical ground state of $V_0(x)$ is zero. Notice that $D_e \gg -c$, which is important as the ground state energy of the nuclear degree of freedom becomes accessible as nuclear kinetic energy within the Ehrenfest approach.\cite{Guo1996, Stock2005}

	In case of the charged molecule, the motion of the nuclear degree of freedom is described by a repulsive generalized Morse potential
	\begin{eqnarray}
		V_d(x) &=&	D_1 \cdot e^{-2\cdot a' (x-x_0')} - D_2 \cdot e^{-a' (x-x_0')}  + V_\infty,
	\end{eqnarray}
	where $D_1=4.52\text{eV}$ and $D_2=0.79\text{eV}$ set the energy scale for the potential, $a'=1.379/\AA$ is the generalized width and $x_0'=1.78\AA$ the position of the generalized minimum. The parameter $V_\infty = -1.5$ eV describes the electron affinity for the dissociated molecule. 
	The choice of $V_\infty$ was motivated by the requirement that the energy of the anti-bonding potential at large distances lies well below the ground state energy of $V_0(x)$. 
	The value of $V_\infty$ used here is to a certain extent arbitrary, but gives representative results. Throughout the paper, we will comment on the importance of $V_\infty$ whenever appropriate.
	
	The shape of the potentials and the parameters are inspired by the model for dissociative electron attachment in CH$_3$Cl in the gas phase.\cite{Fabrikant1991, Fabrikant1994} 
	Investigations of dissociative electron attachment in H$_2$ \cite{Gertitschke1993} and CF$_3$Cl \cite{Hahndorf1994, Wilde1999} yield parameters in the same range.
	We want to emphasize, though, that the goal of our study is to understand the basic mechanisms of current-induced bond rupture for a generic model rather then describing a specific molecule.

	Upon dissociation of the side group of the molecular bridge, the conductance of the junction will change. For example, the bond rupture may destroy the $pi$-conjugation of the molecular backbone resulting in a decrease of the conductance upon dissociation. Within our model, the dependence of the conductance on the nuclear distance is described by the molecule-lead coupling $V_{K}(x)$. To model the mentioned scenario, we use a molecule-lead coupling of the form (see Fig.\ \ref{fig:pots_overview})
	\begin{eqnarray}
		V_{K}(x)    &=&    \overline V_{K} \cdot \left( \frac{1-q}{2} \left[ 1-\tanh\left(\frac{x-\tilde x}{\tilde a} \right) \right] + q \right). \ \label{eq:def_mol_lead_coupling_strength}
	\end{eqnarray}
	Here, $\overline V_{K}$ is the maximal coupling strength between the molecule and the leads. The parameter
	$q=0.05$ determines the coupling strength for large distances, that is $V_{K}(x\rightarrow\infty) = q \overline V_{K}$.
	The distance around which the drop in the molecule-lead coupling occurs is given by $\tilde x=3.5 \AA$, while $\tilde a=0.5\AA$ regulates the width of the region of change.
	
	In the calculations reported below, we assume that both leads have the same temperature $T=300$ K and that the bias voltage, defined as the difference between the chemical potentials $\mu_{\text{L}}$ and $\mu_{\text{R}}$, drops symmetrically such that $\mu_{\text{L}} = -\mu_{\text{R}}$. Inspired by the C-Cl bond, we set the reduced mass to $m=10.54$u.

	In all simulations we assume that the total density matrix factorizes at $t=0$, which describes the scenario that the contact between the previously well separated molecule and the leads is established at $t=0$. 
	Furthermore, the molecule is initially assumed to be in the neutral state, corresponding to a stable equilibrium of the nuclear degree of freedom. Test calculations show that an initial population of the charged state of the molecule can effect the long-time behavior of the systems depending on the molecule-lead coupling strength $\Gamma$, 
	where $\Gamma$ denotes the maximal value in accordance with Eqs.\ (\ref{eq:def_Gamma}) and (\ref{eq:def_mol_lead_coupling_strength}).
	For large $\Gamma > \hbar\omega$, the specific initial electronic state has little influence, as the electronic relaxation via the coupling to the leads is faster than the nuclear response to the initial population.
	For small molecule-lead couplings $\Gamma < \hbar\omega$, however, the initial electronic state is of profound importance as the reaction of the nuclear configuration to the electronic population is faster than the electronic relaxation via the leads. Starting initially in the charged state of the molecule, the molecule always dissociates for $\Gamma \ll \hbar\omega$ because the dissociation process occurs before the molecule stabilizes upon electron detachment to the leads.
	
	For all data presented in the following, we have tested the convergence of the observables with respect to the number of trajectories used for phase-space sampling and the number of poles used to represent the Fermi function in the leads.

\subsection{Coupling to a single lead}\label{sec:gate-voltage}

\begin{figure}[htb!]
		\includegraphics[width = 0.3\textwidth]{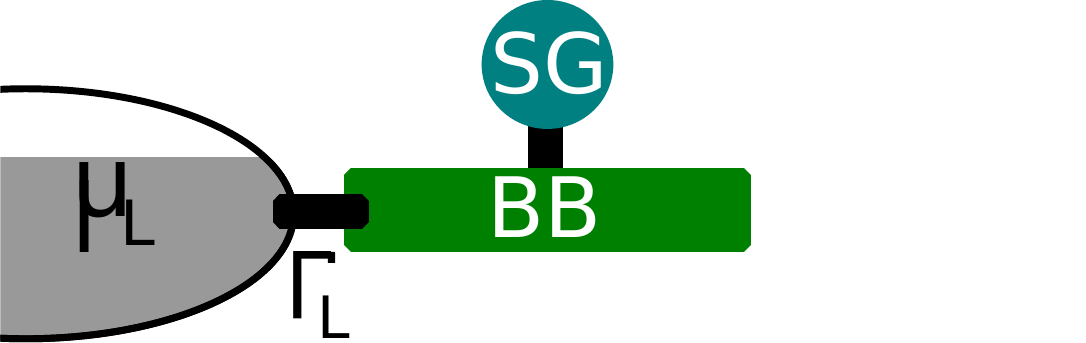}
		\caption{Sketch of the system investigated in Sec.\ \ref{sec:gate-voltage}.}
		\label{fig:sketch_gate_voltage}
	\end{figure}

	We first consider the model system attached to a single lead as depicted in Fig.\ \ref{fig:sketch_gate_voltage}. 
	This setup corresponds to a molecule on a metal surface, which is of interest for studying surface reactions such as desorption or dissociation.\cite{Gao1995_1, Gao1995_2, Brandbyge1995, Saalfrank1996, Vondrak1999}
	In the present context, this coupling scenario serves as starting point, used to introduce the concepts necessary to understand the basic mechanisms of the transport problem. 

	In the case of a single lead (labeled L), the system will assume the equilibrium state provided by the lead in the long-time limit and there is no steady-state current.
	As the chemical potential of the lead $\mu_{\text{L}}$ increases, the energies of the electronic states in the lead are shifted upwards by $\mu_{\text{L}}$. Consequently, the energy of the electrons provided by the lead increases.

	\begin{figure}[htb!]
		\includegraphics[width=0.95\linewidth]{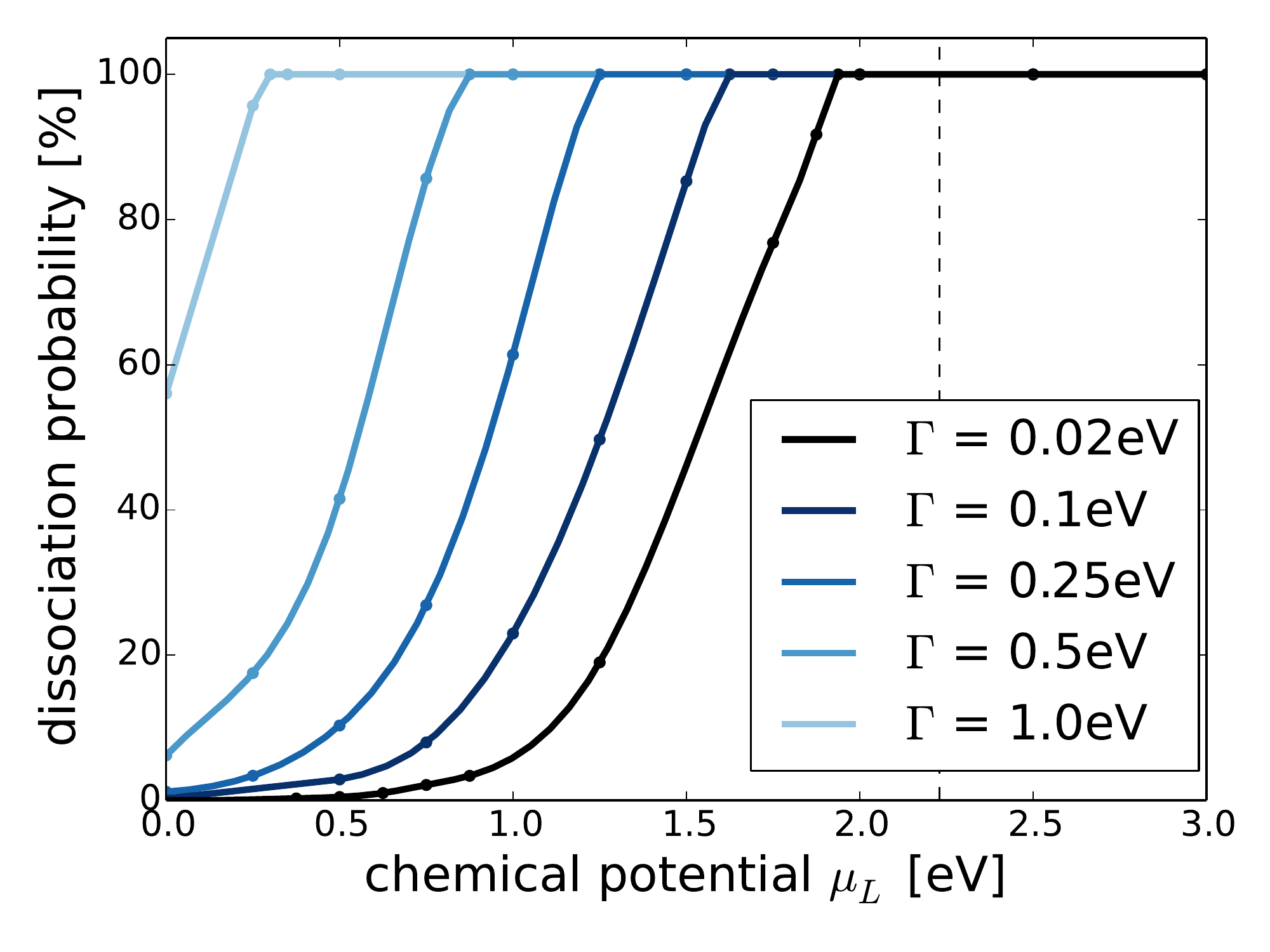}
		\includegraphics[width=0.95\linewidth]{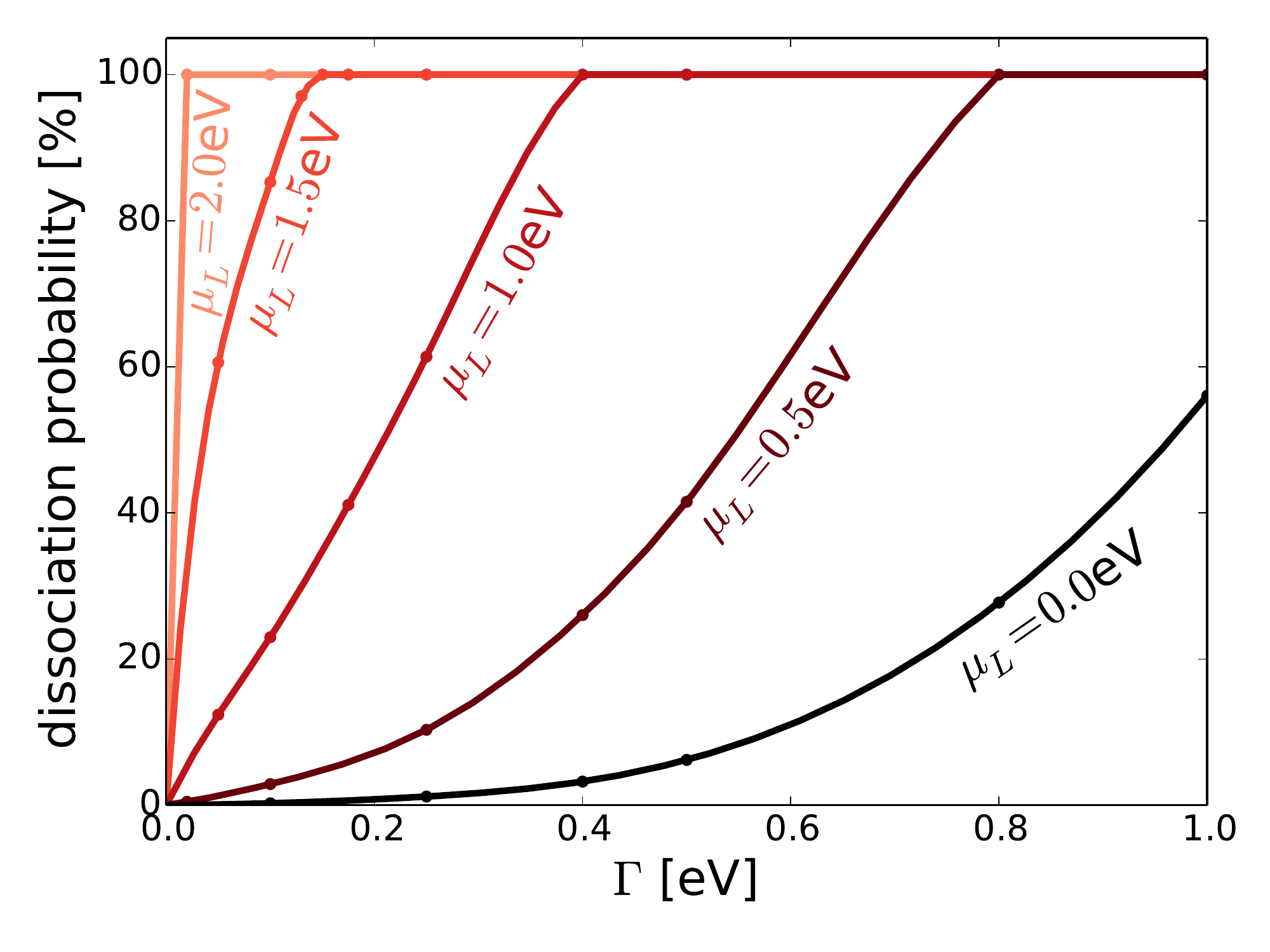}
		\caption{Long-time dissociation probability for the model system attached to a single lead as a function of chemical potential $\mu_{\text{L}}$ for different values of $\Gamma$ (top) and as a function of molecule-lead coupling strength $\Gamma$ for different values of $\mu_{\text{L}}$ (bottom). The points in the plots mark the actual data, the lines serve as a guide for the eye.}
		\label{fig:data_one_lead_only__diss_coupling}
	\end{figure}

	\begin{figure}[htb!]
		\includegraphics[width=\linewidth]{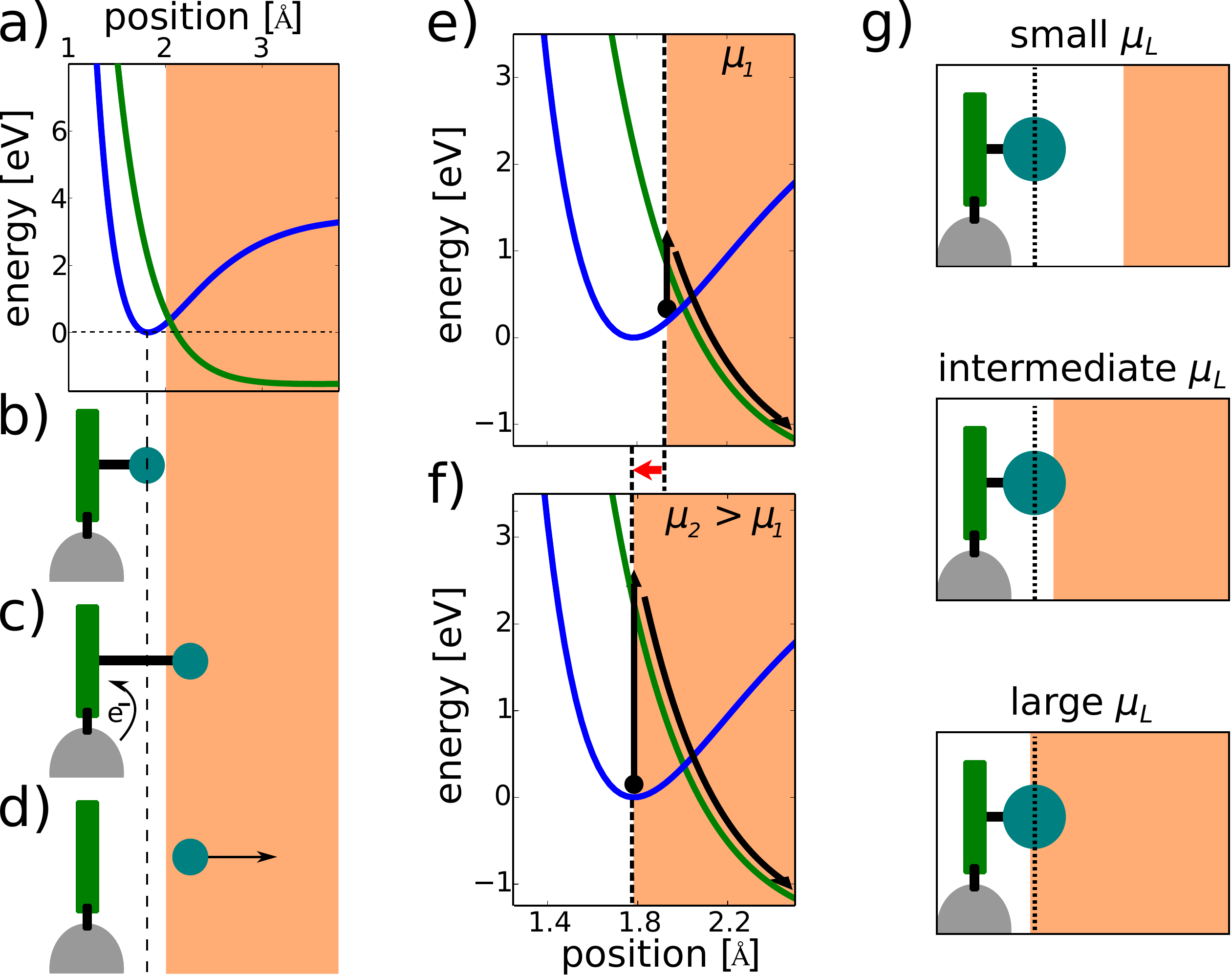}
		\caption{
		 Visualization of the concepts introduced for understanding the dissociation behavior of the model system attached to a single lead.
		  a): Potential energy surfaces of the model system. The orange-shaded area highlights 
		  the {populated regime} whereas the white area marks the {unpopulated regime} for $\mu_{\text{L}}=0$.
		  b) -- d): Representation of the bond rupture process.
		  e) -- g): Illustration of the mechanism by which an increase in bias voltage influences the 
		  extent of the {populated regime}.
		}
		\label{fig:data_one_lead_only__explanation}
	\end{figure}

	The simulated long time-dissociation probability for this model system is depicted in Fig.\ \ref{fig:data_one_lead_only__diss_coupling} as a function of $\mu_{\text{L}}$ (top panel) and as a function of molecule-lead coupling strength (bottom), respectively. 
	We first consider the dissociation probability as a function of $\mu_{\text{L}}$. According to physical intuition it is expected that the system will dissociate above a certain bias voltage, whereas it will be rather stable below this voltage. This behavior is revealed by the simulation results in Fig.\ \ref{fig:data_one_lead_only__diss_coupling} (top).
	The dissociation probability increases monotonically with $\mu_{\text{L}}$ and grows steeply around a certain chemical potential from low dissociation probabilities to $100\%$, giving rise to a threshold like behavior.
	Additionally, a stronger molecule-lead coupling strength $\Gamma$ always leads to an enhanced dissociation probability and reduces the bias for dissociation. That the dissociation probability depends in a non-trivial way on the molecule-lead coupling strength can be seen in Fig.\ \ref{fig:data_one_lead_only__diss_coupling}
	(bottom).

	Dissociation occurs only if the molecular electronic state is populated, i.e.\ the molecule is charged, implying that the nuclear motion is governed by the anti-bonding potential $V_d(x)$. 
	Thus the data can be rationalized based on the different charge states of the molecule.
	For fixed nuclei, whether an electron from the lead can populate the molecule depends on the energy-difference between $V_d(x)$ and $V_0(x)$, which is a function of the nuclear coordinate $x$. For $V_d(x)-V_0(x) < \mu_{\text{L}}$, an electron can be transferred from the leads to the molecule and, as a result, the molecule is charged. 
	In the following, we will refer to the corresponding $x$-values as the \textit{populated regime}.
	The range of $x$-values where the molecule can not be populated by an electron from the lead, $V_d(x)-V_0(x) > \mu_{\text{L}}$, will be termed \textit{unpopulated regime}.
	Fig.\ \ref{fig:data_one_lead_only__explanation}a depicts this relation for $\mu_{\text{L}}=0$, where the orange-shaded area beyond the point where $V_d(x) = V_0(x)$ corresponds to the {populated regime}, i.e.\ the nuclear coordinates where the molecule can be populated by an extra electron from the lead, and the white area highlights the {unpopulated regime}. 
	As the $x$-value satisfying the condition $V_d(x)-V_0(x) = \mu_{\text{L}}$ separates the {populated regime} from the {unpopulated regime}, the absolute value of $V_d(x)$ and consequently $V_\infty$ is important for the extent of the two regimes. 
	As $V_\infty$ enters $V_d(x)$ as an additive constant, a change in $V_\infty$ has the same influence on the dissociation probability as a change of $\mu_{\text{L}}$ by the same amount, as is apparent from the equation $V_d(x)-V_0(x) = \mu_{\text{L}}$.

	The strict separation of possible values for the nuclear coordinate in terms of electronic population is only a qualitative criterion, which neglects broadening effects leading to partial electronic population and the influence of the dynamics of the nuclear degree of freedom. Both effects are accounted for in the simulations. Their influence on the dissociation probability can be understood in the following way.
	In the approach used in this paper, the nuclear dynamics is described by a set of trajectories with different initial conditions, representing the non-localized nature of the nuclear degree of freedom. The most probable location for the nuclei is close to the minimum of $V_0(x)$ as depicted in Fig.\ \ref{fig:data_one_lead_only__explanation}b. However, there is a non-vanishing probability for the nuclear degree of freedom to be located at larger $x$-values within the {populated regime} as shown in Fig.\ \ref{fig:data_one_lead_only__explanation}c. If the nuclear coordinate reaches the {populated regime} and stays in this regime an amount of time, which is sufficient for the electrons to populate the molecule (a time-scale given by $\Gamma$), the molecule will dissociate, as is depicted in Fig.\ \ref{fig:data_one_lead_only__explanation}d. The fact that $\Gamma$ sets the time-scale for electrons populating the molecule and that it leads to a broadening of the electronic level, thus smearing the border between the {populated} and the {unpopulated regime}, leads to the nontrivial relation between the dissociation probability and $\Gamma$ depicted in Fig.\ \ref{fig:data_one_lead_only__diss_coupling} (bottom).

	Next, we consider the threshold-like behavior of the dissociation probability as a function of bias voltage in Fig.\ \ref{fig:data_one_lead_only__diss_coupling} (top). 
	A varying chemical potential $\mu_{\text{L}}$ of the lead influences the extent of the {populated} and {unpopulated regime}.
	This is depicted in Figs.\ \ref{fig:data_one_lead_only__explanation}e and f, where the vertical black arrow indicates the energy that needs to be provided by the lead in order to populate the molecular electronic state. 
	For low bias voltages, the nuclear coordinate must deviate strongly from the equilibrium position in order to dissociate, resulting in a low dissociation probability (Fig.\  \ref{fig:data_one_lead_only__explanation}g top). 
	Upon increasing $\mu_{\text{L}}$, the {populated regime} also includes smaller $x$-values (Figs.\ \ref{fig:data_one_lead_only__explanation}e and f), such that the deviation from the nuclear equilibrium position necessary for dissociation diminishes. 
	The threshold-like increases in dissociation probability then occurs around values of $\mu_{\text{L}}$, where the nuclear equilibrium position enters the populated regime (Fig.\ \ref{fig:data_one_lead_only__explanation}g bottom).

	Fig.\ \ref{fig:data_one_lead_only__diss_coupling} (top) indicates that the threshold for dissociation decreases with $\Gamma$ and that it is always lower than the classical expectation for the threshold $\mu_{\text{L}} = V_d(x_0)-V_0(x_0) = 2.33$ eV (vertical dashed black line in Fig.\ \ref{fig:data_one_lead_only__diss_coupling} (top)).
	The main reason for this shift towards lower $\mu_{\text{L}}$ and the dependence on $\Gamma$ is the non-zero population of the molecular electronic state due to the broadening by molecule-lead coupling. 
	The partial population of the anti-bonding state pushes the nuclear equilibrium position outwards, facilitating dissociation at lower $\mu_{\text{L}}$.
	This effect is further enhanced by thermal broadening and the initial dynamics induced upon establishing the contact between the molecule and the lead at $t=0$.

	\begin{figure}[htb!]
		\includegraphics[width=1.1\linewidth]{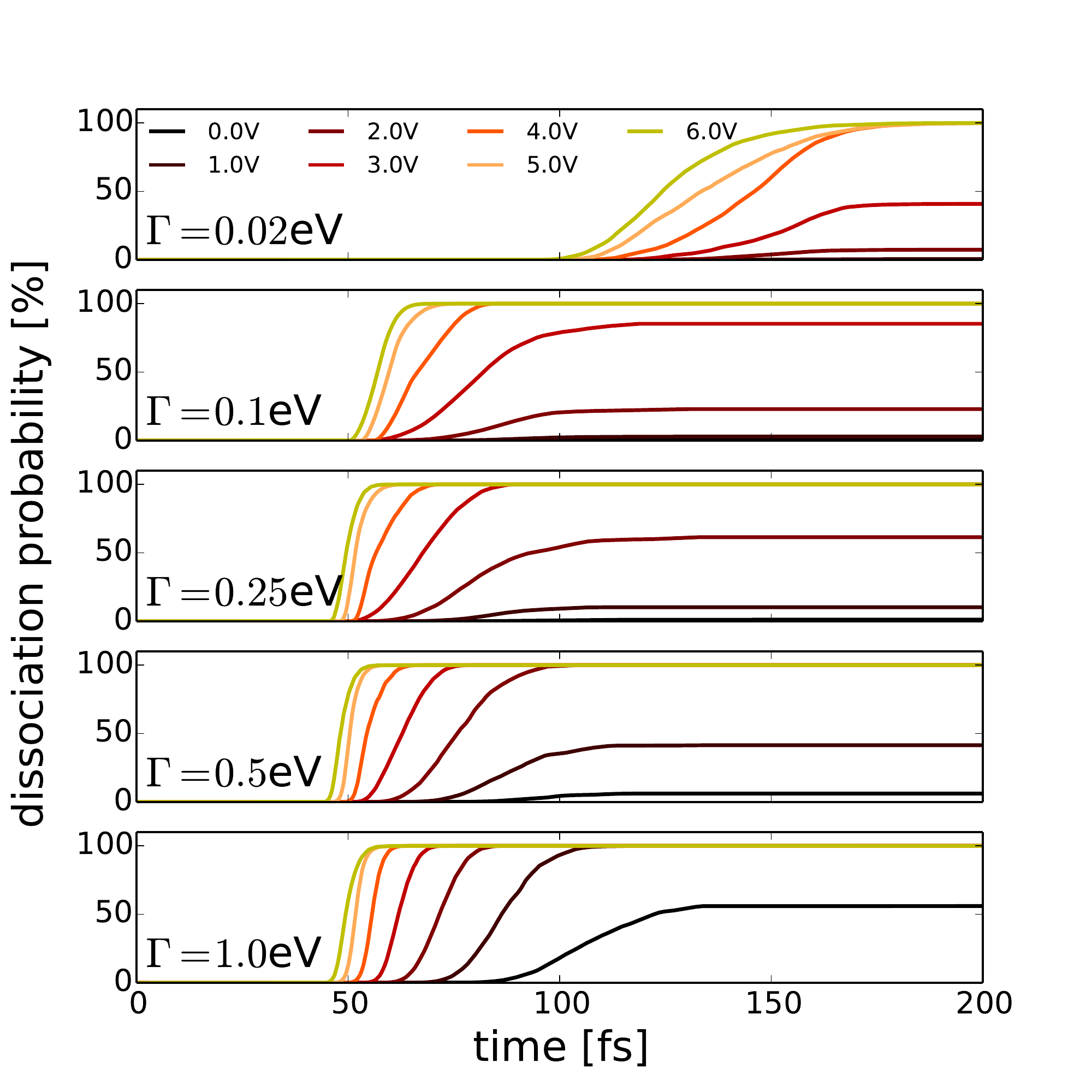}
		\caption{Dissociation probability as a function of time for the model system attached to a single lead. The lines correspond to different chemical potentials $\mu_\text{L}$. The molecule-lead coupling strength $\Gamma$ is increased from the top to the bottom panel.}
		\label{fig:data_one_lead_only__diss_time}
	\end{figure}

	Fig.\ \ref{fig:data_one_lead_only__diss_time} shows the dissociation probability as a function of time. The final value of the dissociation probability is reached rather quickly after the contact between the molecule and the lead is established, within a time scale of $100$ fs.
	This dissociation time decreases moderately upon increasing bias voltage for any value of $\Gamma$. This in line with the interpretation that the extent of the {populated regime} increases towards smaller $x$-values with increasing $\mu_{\text{L}}$ (Figs.\ \ref{fig:data_one_lead_only__explanation}e--g).
	For $\Gamma=0.1$ eV -- $1.0$ eV, the dissociation time shows little dependence on the molecule-lead coupling strength. 
	For $\Gamma=0.02$ eV, the dissociation time increases as only in this case the electron dynamics (a time-scale set by $\Gamma$) is slower than the nuclear motion (a time-scale set by $\hbar\omega$).

\subsection{Symmetric molecule-lead coupling scenario}\label{sec:symmetric}

\begin{figure}[htb!]
		\includegraphics[width = 0.3\textwidth]{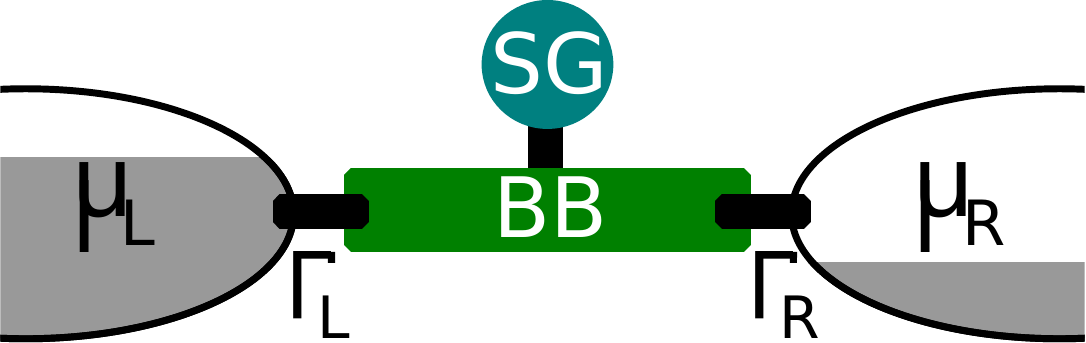}
		\caption{Sketch of the system investigated in Sec.\ \ref{sec:symmetric}. The coupling to both leads is identical, $\Gamma_\text{L}(x) = \Gamma_\text{R}(x)$.}
		\label{fig:sketch_symm}
	\end{figure}

	In the remainder of this paper, we study model systems attached to two leads, which describe the scenario of a molecular junctions. 
	For a finite bias voltage, these systems approach a nonequilibrium  steady state in the long-time limit with a finite current.
	In this section, we analyze the scenario where the molecule couples with the same strength to the left and to the right lead, $\Gamma_\text{L}(x) = \Gamma_\text{R}(x)$, as depicted in Fig.\ \ref{fig:sketch_symm}. 
	As in Sec.\ \ref{sec:gate-voltage}, $\Gamma$ denotes the maximal coupling strength which is identical for both leads.

	\begin{figure}[ht!]
		\includegraphics[width=0.95\linewidth]{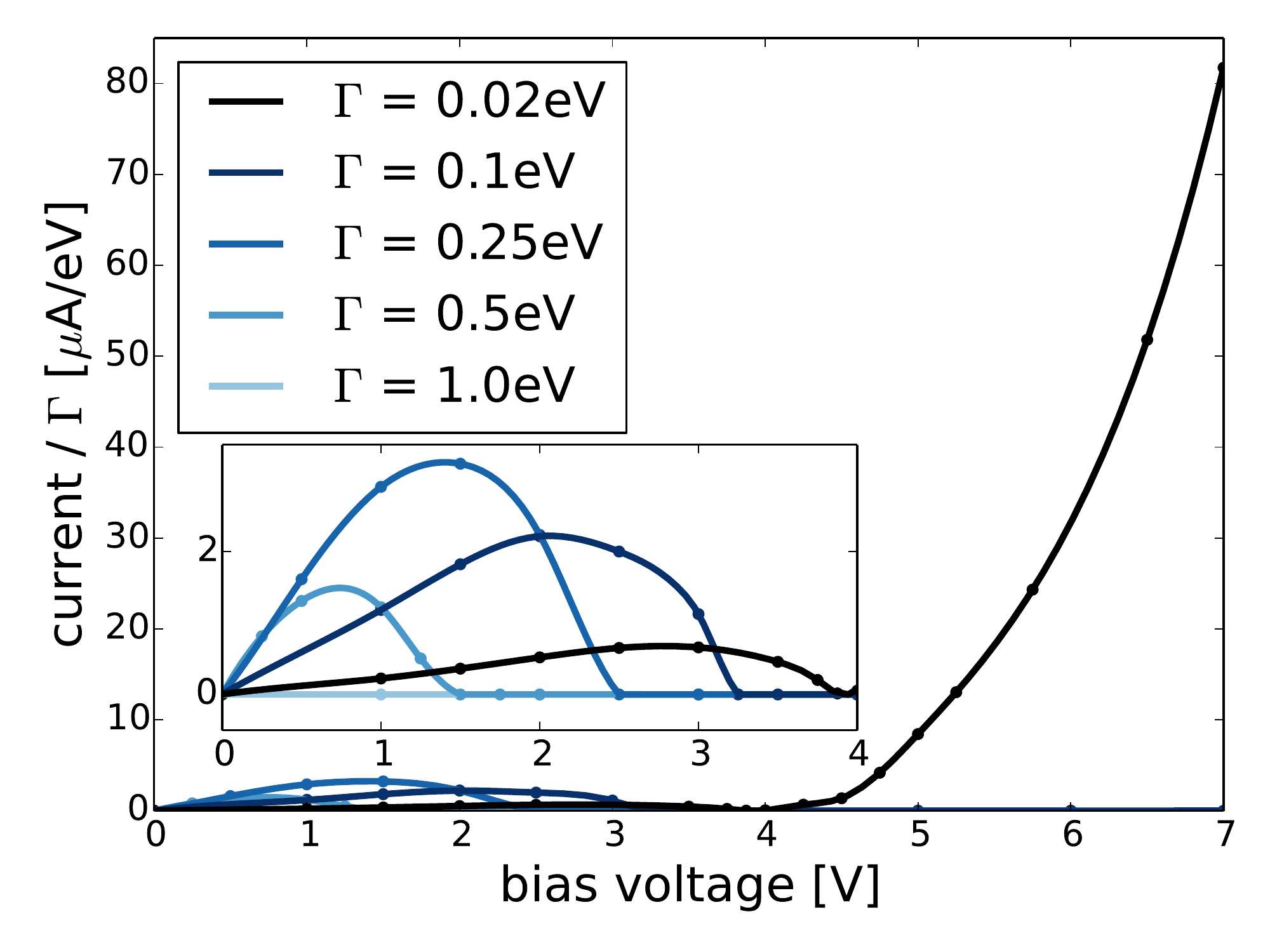}
		\includegraphics[width=0.95\linewidth]{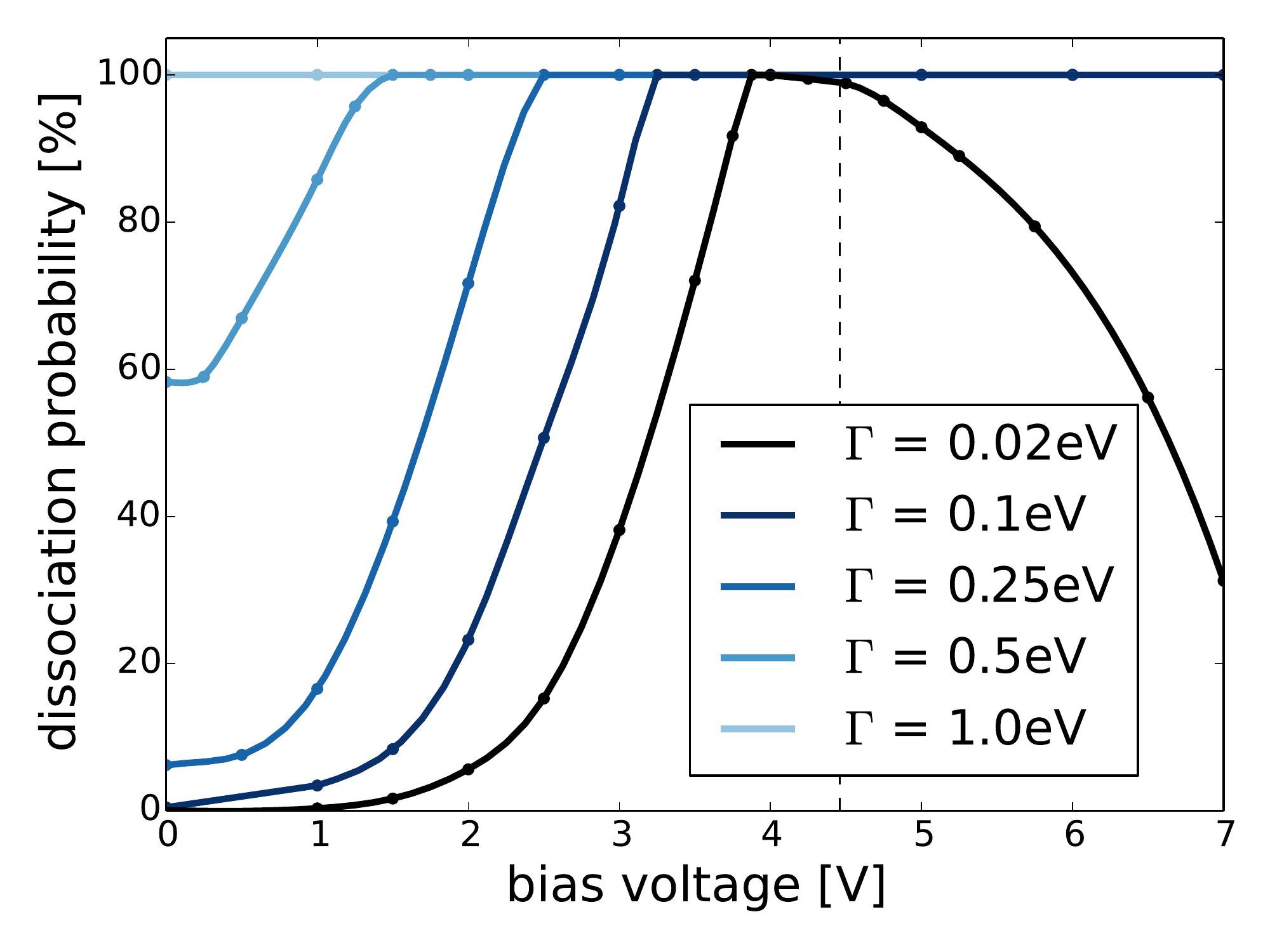}
		\includegraphics[width=0.95\linewidth]{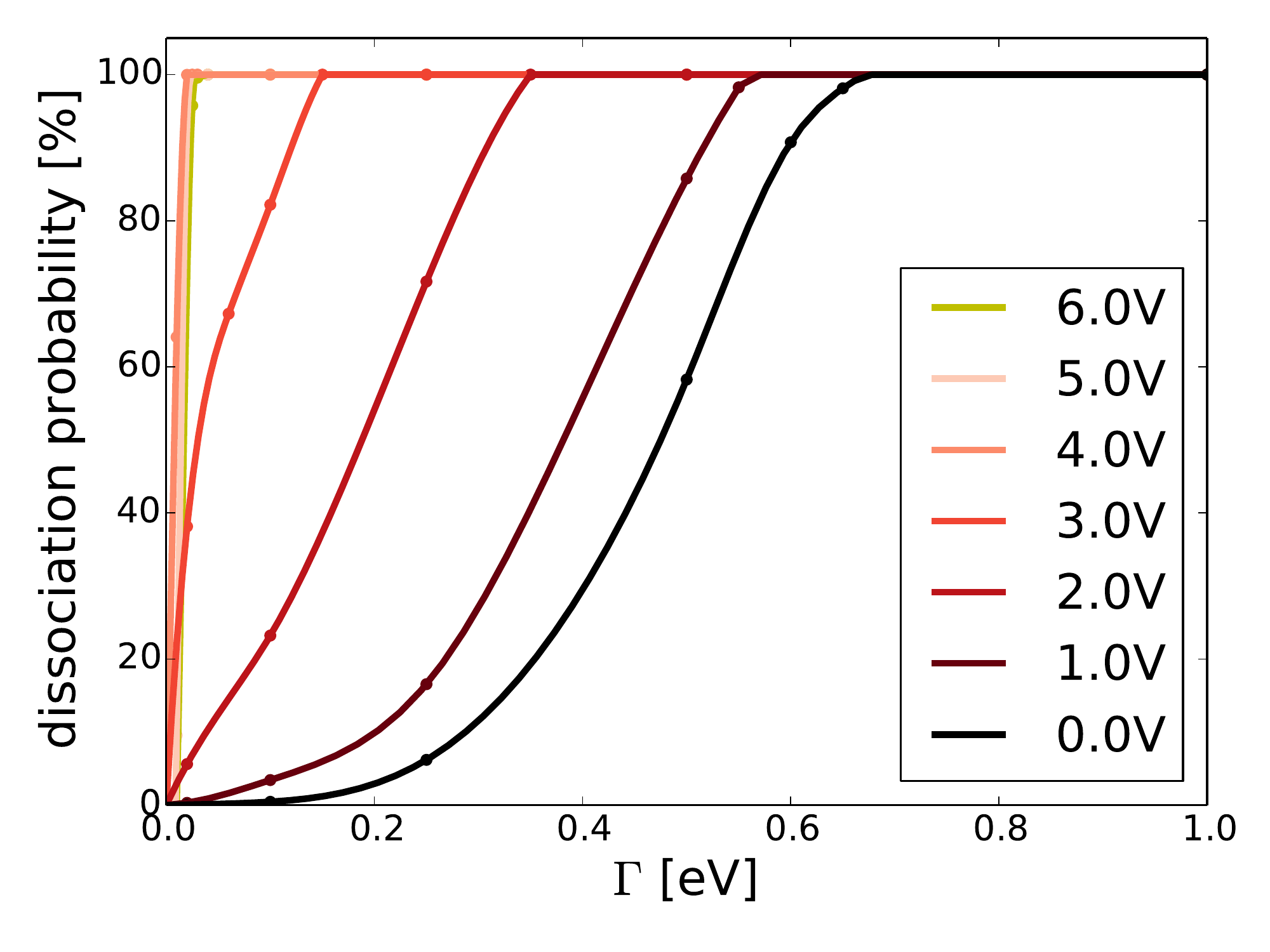}
 		\caption{Average current normalized by $\Gamma$ for the model system with $\Gamma_\text{L}(x) = \Gamma_\text{R}(x)$ as a function of bias voltage for different values of molecule-lead coupling $\Gamma$ (top).
 		Long-time dissociation probability as a function of applied bias voltage for different values of $\Gamma$ (middle). Notice that for bias voltages above $3.25$ V, the dissociation probability for $\Gamma=0.1$ eV -- $1.0$ eV is $100\%$ such that the corresponding lines are on top of each other.
 		Dissociation probability as a function of $\Gamma$ for different voltages (bottom). The points in the plots mark the actual data, the lines serve as a guide for the eye.}
		\label{fig:data_symm__diss_Gamma}
	\end{figure}
	
	The average current for this model normalized by $\Gamma$ is shown in Fig.\ \ref{fig:data_symm__diss_Gamma} (top) as a function of applied bias.
	The corresponding long-time dissociation probability is depicted in Figs.\ \ref{fig:data_symm__diss_Gamma} as a function of applied bias voltage (middle) and as a function of molecule-lead coupling strength (bottom).	
	We first focus on the dissociation probability.
	The results in Fig.\ \ref{fig:data_symm__diss_Gamma} (middle) exhibit the threshold-like onset of dissociation already known from the single lead case in Sec.\ \ref{sec:gate-voltage}. However, the dissociation probability rises slower close to zero bias compared to the single lead setup. Most remarkably, the dissociation probability decreases for bias voltages above $4.5$ V with increasing bias for the weak coupling case $\Gamma=0.02$ eV. 

	\begin{figure*}[htb!]
		\includegraphics[width = 0.99\linewidth]{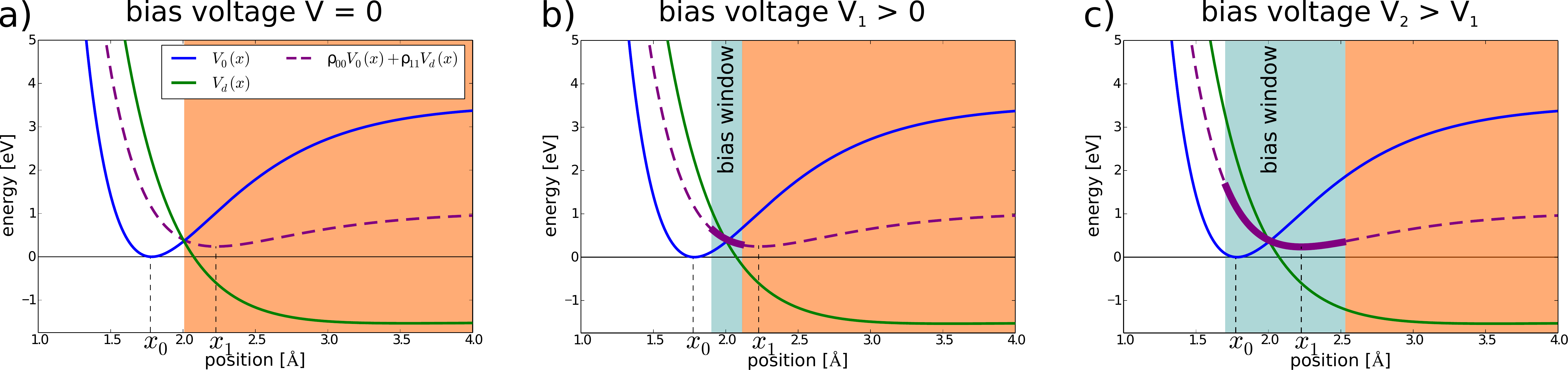}
		\caption{Potential energy surfaces and average potential $V_{\text{av}}(x)= \rho_{00} V_0(x)+ \rho_{11} V_d(x)$ for the coupling scenario $\Gamma_\text{L}(x) = \Gamma_\text{R}(x)$ for three exemplary bias voltages $0 < V_1 < V_2$. The white area represents the {unpopulated regime}, the orange area is the {populated regime}, the blue shaded area highlights the {conducting regime}. The individual pictures a) to c) represent different applied bias voltages. As the bias is increased, the extent of the {conducting regime} enlarges and so is the zone of influence of the average potential. In picture c) the minimum $x_1$ of $V_{\text{av}}(x)$ enters the {conducting regime} thus allowing for a stable conducting molecule.}
		\label{fig:symm_explanation_voltage_window}
	\end{figure*}

	In order to explain this, we consider again a partitioning of the range of possible $x$-values according to the population of the molecular electronic state. 
	If attached to two leads, the molecule allows for resonant transport if the electronic state lies within the bias window, that is $\mu_{\text{L}} > V_d(x)-V_0(x) > \mu_{\text{R}}$, resulting in a partially populated molecular electronic state.
	Consequently, there are three possible charge states of the molecule (populated, unpopulated, partially populated), corresponding to the three regimes of nuclear coordinates $x$, namely the \textit{populated}, the \textit{unpopulated}, and the \textit{conducting regime}, which are highlighted in Fig.\ \ref{fig:symm_explanation_voltage_window} by different colors.
	In order to include the nuclear motion for a molecule in the resonant transport regime into the consideration, we introduce an average potential energy surface $V_{\text{av}}(x) = \rho_{00} V_0(x) + \rho_{11}V_d(x)$, which is depicted as a purple line in Fig.\ \ref{fig:symm_explanation_voltage_window}. There is an optimal nuclear position, $x_1$, for the molecule under current, given by the $x$-value minimizing $V_{\text{av}}(x)$, with $x_1 > x_0$.
	Notice that for the scenario $\Gamma_\text{L}(x) = \Gamma_\text{R}(x)$ studied in this section, the current across the molecule will eventually lead to an electronic state that is about half populated, i.e.\
	$V_{\text{av}}(x) \approx (V_0(x) + V_d(x))/2$.  

	Fig.\ \ref{fig:symm_explanation_voltage_window} visualizes the three regimes and $V_{\text{av}}(x)$ for different applied bias voltages. 
	As the bias is increased, the voltage window corresponding to the conducting regime opens up, pushing the {populated regime} outwards.
	For small bias voltages as in Fig.\ \ref{fig:symm_explanation_voltage_window}b, there is a small {conducting regime} around the position $V_d(x)=V_0(x)$, however the minimum $x_1$ of $V_{\text{av}}(x)$ lies outside this regime. As a result, the resonant transport pushes the nuclei to larger distances, i.e.\ towards dissociation. Because $V_{\text{av}}(x)$ is less steep than $V_d(x)$  within the {conducting regime}, the force exerted on the nuclei is smaller compared to the one lead case, explaining the less steep increase in dissociation probability for low bias voltages in Fig.\ \ref{fig:data_symm__diss_Gamma} (middle). Upon increasing the bias voltage, the {conducting regime} increases (see Fig.\ \ref{fig:symm_explanation_voltage_window}c). Above a certain bias voltage, also the minimum $x_1$ of $V_{\text{av}}(x)$ lies within this regime. As a consequence, the molecule under current can be stable, resulting in the decrease of dissociation probability for high bias voltages for a molecule-lead coupling strength $\Gamma=0.02$ eV in Fig.\ \ref{fig:data_symm__diss_Gamma} (middle). 
	The decrease in dissociation probability upon increase of the applied bias thus occurs when the minimum $x_1$ of $V_{\text{av}}(x)$ enters the {conducting regime}. Note that the location of the {conducting regime} in nuclear coordinate space is given by the energy difference between $V_d(x)$ and $V_0(x)$ and, therefore, depends on $V_\infty$.

\begin{figure}[htb!]
		\includegraphics[width = \linewidth]{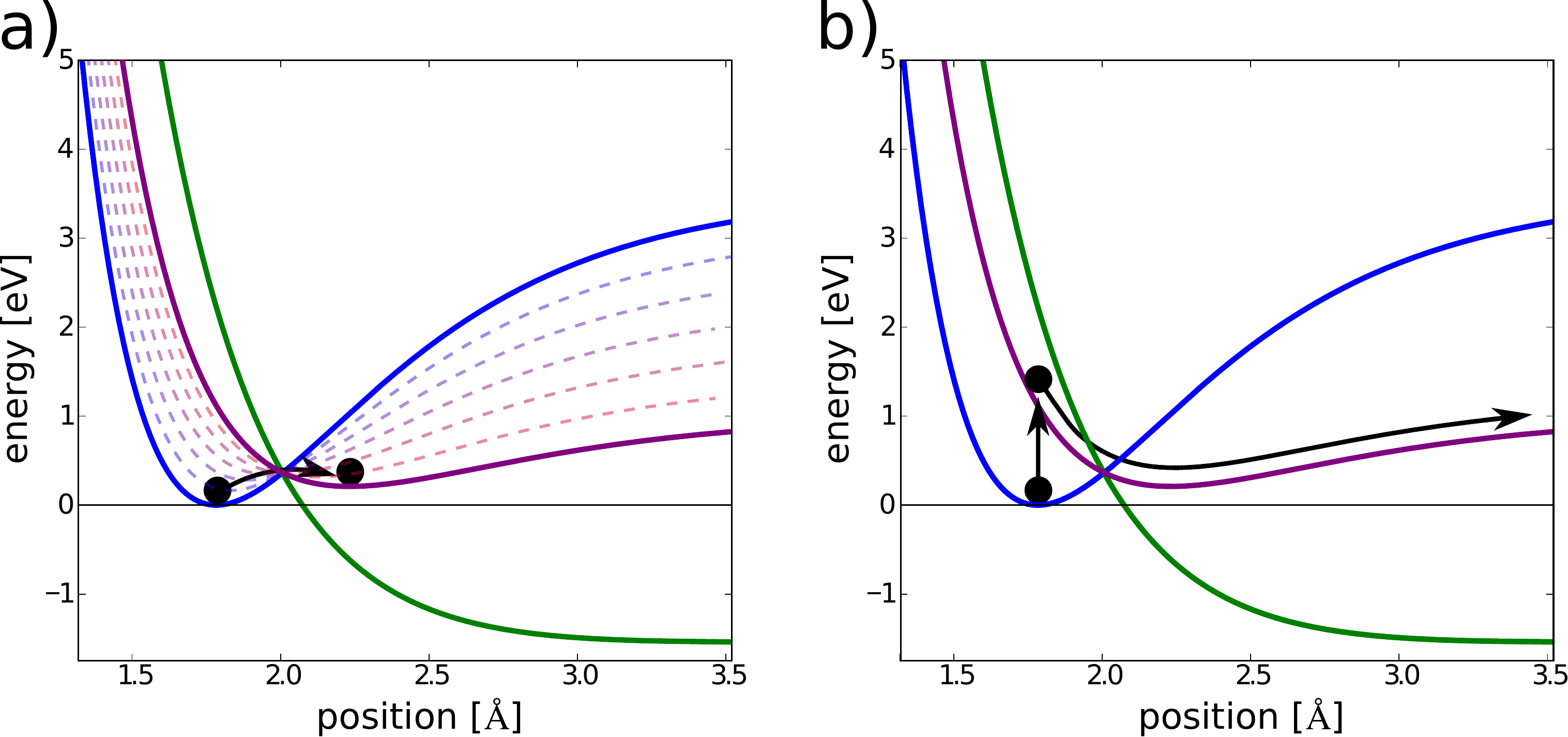}
		\caption{Influence of molecule-lead coupling $\Gamma$ on the dissociation probability. 
		a): For $\hbar\omega > \Gamma$, the nucleus moves in a slowly varying potential, thus assuming a stable nuclear configuration.
		b): For $\hbar\omega < \Gamma$, the nucleus instantly moves on $V_{\text{av}}(x)$, resulting in dissociation.
		}
		\label{fig:sketch_adiabatic_VS_anti-adiabatic}
	\end{figure}

	The reason for the finding that there is a decrease in dissociation probability with bias only for $\Gamma=0.02$ eV is the different time-scales for electronic motion, which is determined by $\Gamma$, and for nuclear motion, which is characterized by $\hbar\omega$. 
	In the anti-adiabatic regime, $\hbar\omega > \Gamma$, where the electrons are slower than the nuclei, the nuclei move within a slowly varying potential energy surface. This situation is depicted in Fig.\ \ref{fig:sketch_adiabatic_VS_anti-adiabatic}a, where the nuclei, constantly located at the minimum of the potential energy surface, is slowly pushed outwards by the increasing population of the electronic state. 
	The dependence of this effect on the molecule-lead coupling  strength can be observed in Fig.\ \ref{fig:data_symm__diss_Gamma} (bottom), where the dissociation probability is displayed as a function of $\Gamma$. For very small $\Gamma \approx 0 - 0.03$ eV, the lines representing the bias voltages $5$ V and $6$ V lie below the line for $4$ V, demonstrating that the effect is only present for small molecule-lead coupling strengths and high bias voltages.
	In the adiabatic regime $\hbar\omega < \Gamma$, the electrons are much faster than the nuclear motion, such that the nuclei move under the force of (quasi-)equilibrated electrons. For high bias voltages, this means that the nuclear motion is instantly governed by $V_{\text{av}}(x)$, once the contact between molecule and leads is established. As $V_{\text{av}}(x_0) > V_{\text{av}}(\infty)$, this results in the dissociation of the bond as depicted in Fig.\ \ref{fig:sketch_adiabatic_VS_anti-adiabatic}b.
	
	Next, we consider the average electronic current across the molecular junction shown in Fig.\ \ref{fig:data_symm__diss_Gamma} (top). 
	For low bias voltages, we observe a small current which rises with bias, corresponding to the non-resonant current across the molecule (see inset Fig.\ \ref{fig:data_symm__diss_Gamma} (top)). Accordingly, the current is higher for larger $\Gamma$.
	At the voltage around which the threshold-like onset of dissociation occurs, the non-resonant current drops significantly. With the onset of dissociation, the probability for the molecule to be in the poorly conducting dissociated state is enhanced resulting in a decrease in the average current. Finally, for $\Gamma=0.02$ eV, in the regime where we observe the decrease of dissociation probability with bias, there is a pronounced current that rises with bias. This is consistent with the explanation above on how the molecule can assume a stable transport configuration. 
	
	As in the one lead case, the dissociation times are on the order of $100$ fs (data not shown). Generally, the dissociation times moderately decrease with increasing bias voltage, only for $\Gamma=0.02$ eV they exhibit a slight increase. Again, the dissociation times are rather insensitive to $\Gamma$.
	
	We close this section with a few comments on related work. The mechanism that the partial occupation of electronic states induced by an electrical current influences molecular bonds was studied before by other authors. For example, \citet{Brandbyge2003}  argued that this effect can strengthen or weaken bonds in a molecule under bias. \citet{Hussein2010} considered a harmonic nuclear mode within the adiabatic approximation. They found that the force (and consequently the potential) depends on the electronic population and that the effective potential exhibits several minima corresponding to different charge state of the molecule. This work was extended beyond the adiabatic approximation by Metelmann and Brandes.\cite{Metelmann2011} \citet{Wilner2014} also considered a single electronic state coupled to a harmonic bath and found that the relaxation dynamics as well as the possibility for bistability is related to the different minima in the potential energy surface for different charge states. \citet{Dzhioev2011} used nonequilibrium, current-depended potential energy surfaces to study  current-induced chemical reactions of the H$_2^+$ molecule. They found that the nonequilibrium correction is due to the variation of the electronic population, which is most significant if the electronic state is within the bias window. Furthermore, \citet{Pozner2014} investigated charge transport in a double quantum dot system and found that the quantum dot distance is associated with the average electronic population, which in turn is influenced by the current.

\subsection{Asymmetric molecule-lead coupling scenario}\label{sec:asymmetric}

\begin{figure}[htb!]
		\includegraphics[width = 0.3\textwidth]{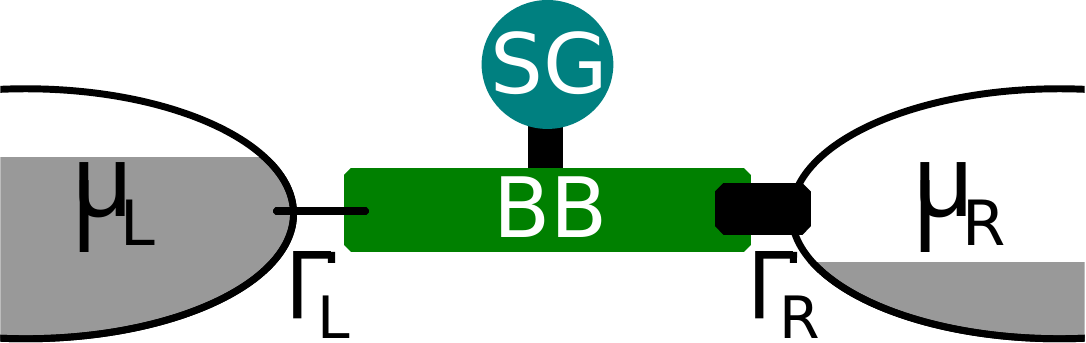}
		\caption{Sketch of the system investigated in Sec.\ \ref{sec:asymmetric}. The coupling to both leads is different, $\Gamma_\text{L}(x) = 0.25 \cdot \Gamma_\text{R}(x)$.}
		\label{fig:sketch_asymm}
	\end{figure}

	In this final section, we consider the case that the system is more strongly coupled to one of the leads. Such an asymmetric coupling scenario can be found in STM experiments, where the molecule is more strongly bound to the substrate than to the STM tip. As an example, we consider the case $\Gamma_\text{L}(x) = 0.25 \cdot \Gamma_\text{R}(x)$, which results in a partial population of the molecular electronic state of about $\rho_{11} \approx 0.2$ for positive bias in the resonant transport regime. The setup is sketched in Fig.\ \ref{fig:sketch_asymm}. 
	
	\begin{figure}[h!]
	\vspace*{-0.5cm}
		\includegraphics[width=0.95\linewidth]{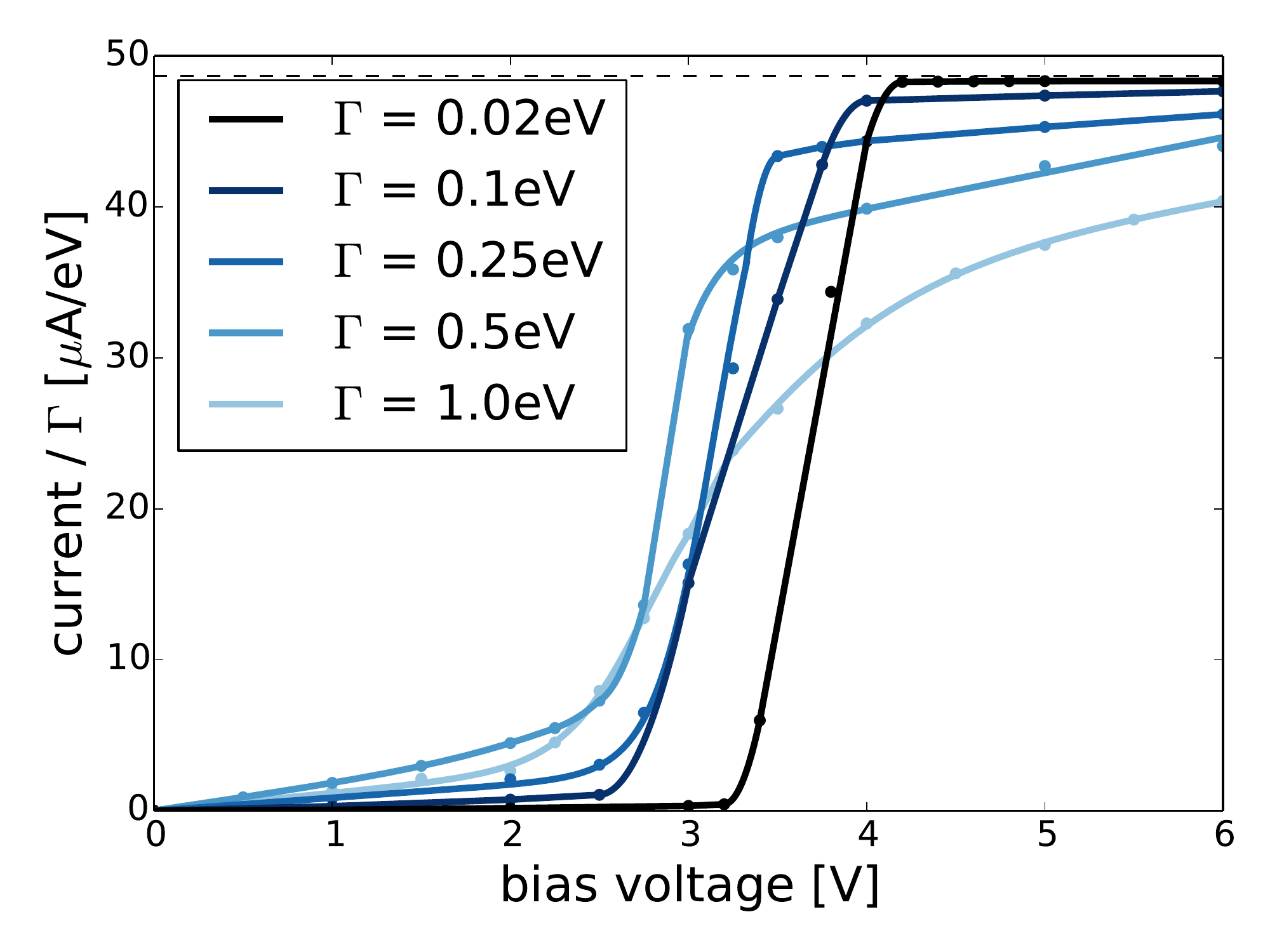}
		\includegraphics[width=0.95\linewidth]{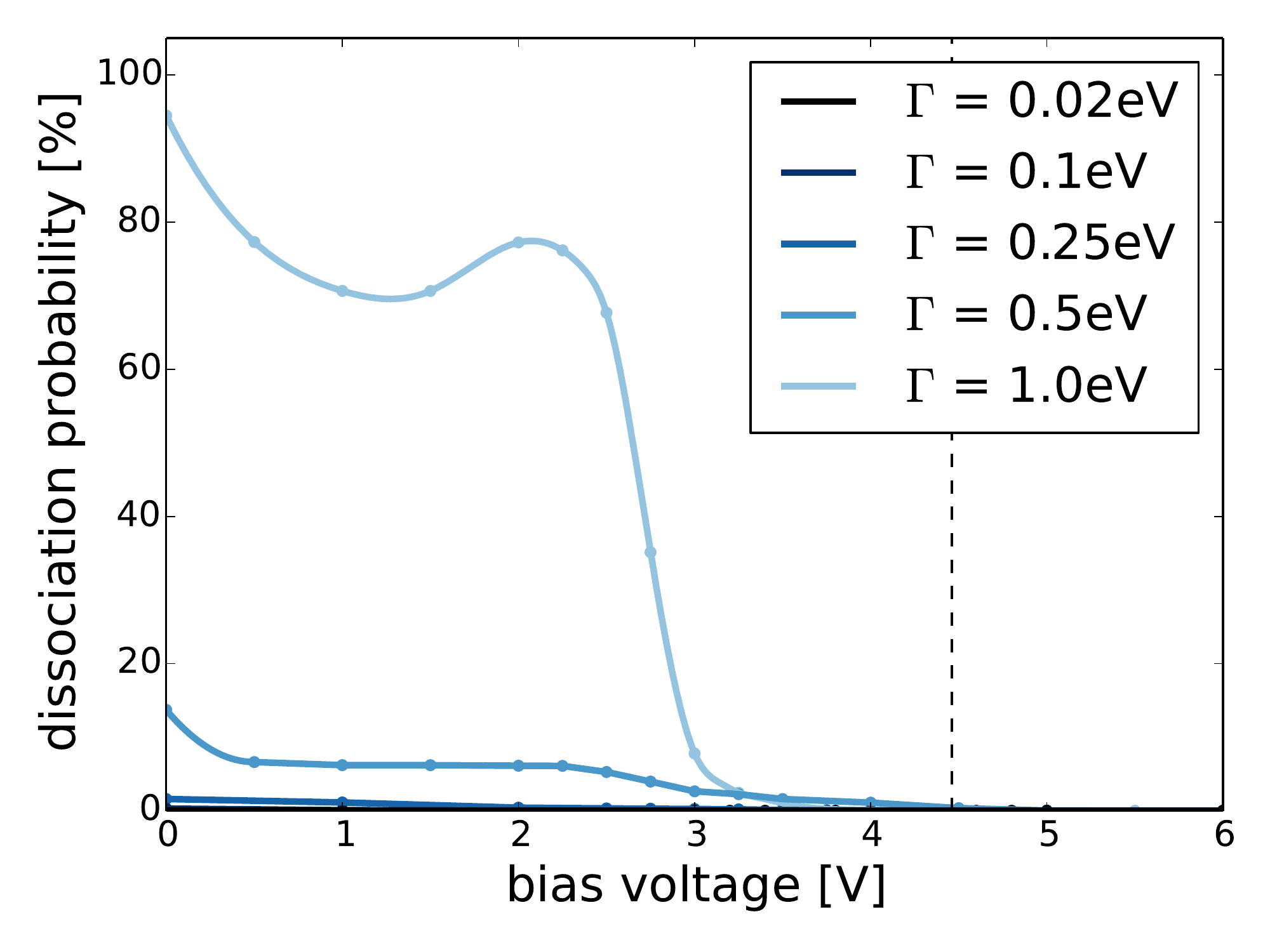}
		\includegraphics[width=0.95\linewidth]{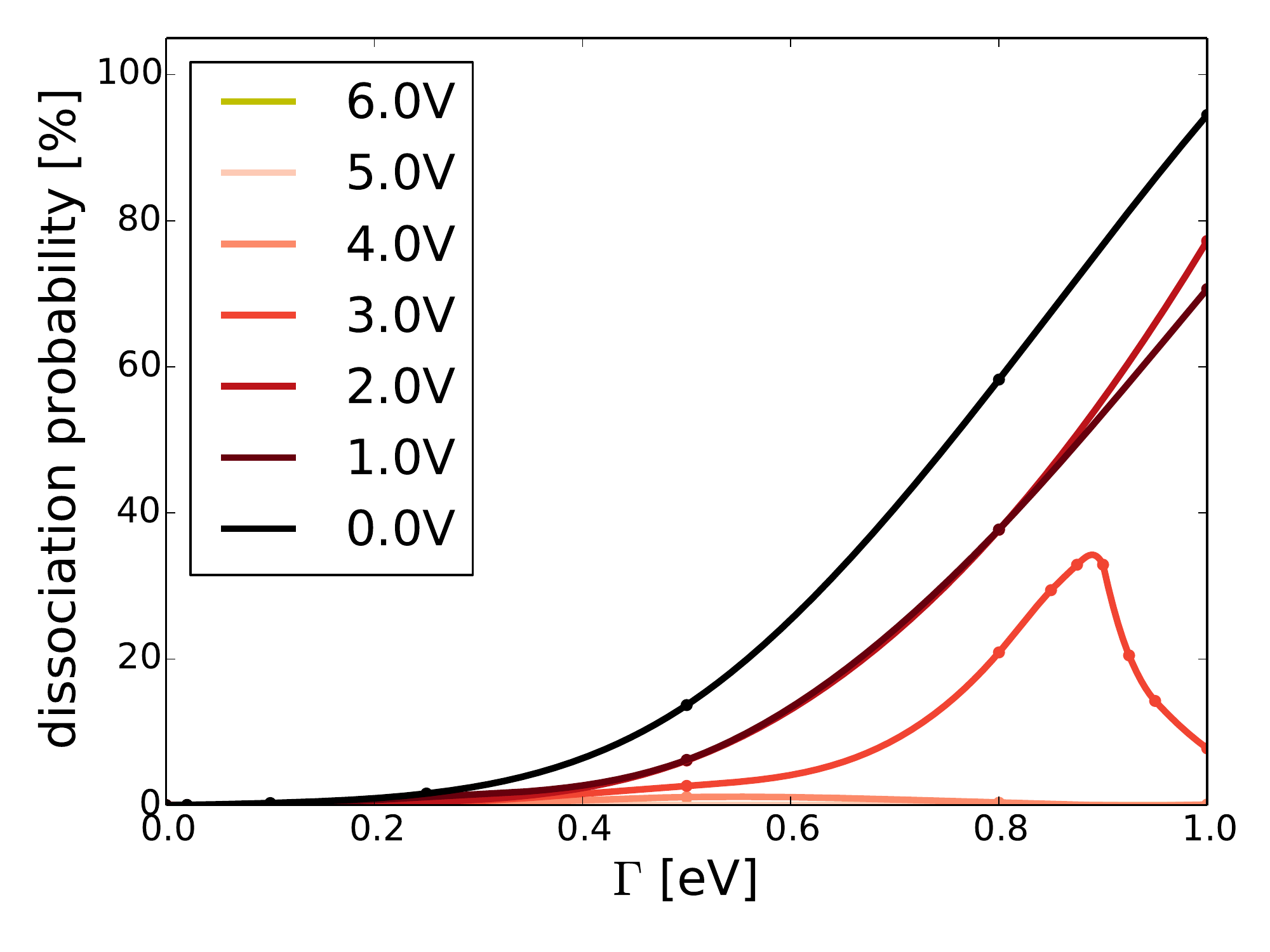}
		\caption{Average current normalized by $\Gamma$ for the model system more strongly coupled to the right lead as a function of applied bias voltage for different values of $\Gamma$ (top).
		The horizontal dashed black line corresponds to the maximal possible current $I_{\text{L/R}}/\Gamma = 0.2 e/\hbar$.
 		The long-time dissociation probability for this system as a function of applied bias voltage for different values of $\Gamma$ (middle) and as a function of molecule-lead coupling strength $\Gamma$ for different voltages (bottom). The points in the plots mark the actual data, the lines serve as a guide for the eye.}
		\label{fig:data_asymm__diss_Gamma}
	\end{figure}
	
	The average current for the asymmetric model normalized by $\Gamma$ is shown in Fig.\ \ref{fig:data_asymm__diss_Gamma} (top) as a function of applied bias.
	The corresponding long-time dissociation probability is depicted in Figs.\ \ref{fig:data_asymm__diss_Gamma} as a function of applied bias voltage (middle) and as a function of molecule-lead coupling strength (bottom), respectively. 
	We first consider the dissociation probability. The results in Fig.\ \ref{fig:data_asymm__diss_Gamma} (middle) show that the dissociation probability always decreases with bias for molecule-lead coupling strengths $\Gamma=0.02$ eV -- $0.5$ eV. For $\Gamma=1.0$ eV, this overall trend is broken by a local maximum in dissociation probability at around $2$ V bias voltage.
	Furthermore, the high-bias dissociation probability is lower for $\Gamma=1.0$ eV than for $\Gamma=0.5$ eV. Fig.\ \ref{fig:data_asymm__diss_Gamma} (bottom) demonstrates that the dissociation probability depends in a nonlinear way on the molecule-lead coupling strength $\Gamma$. Particularly striking is the result for a bias voltage of $3$ V, which shows a pronounced peak structure at about $\Gamma\approx0.9$ eV.
	
	\begin{figure*}[htb!]
		\includegraphics[width = 0.99\linewidth]{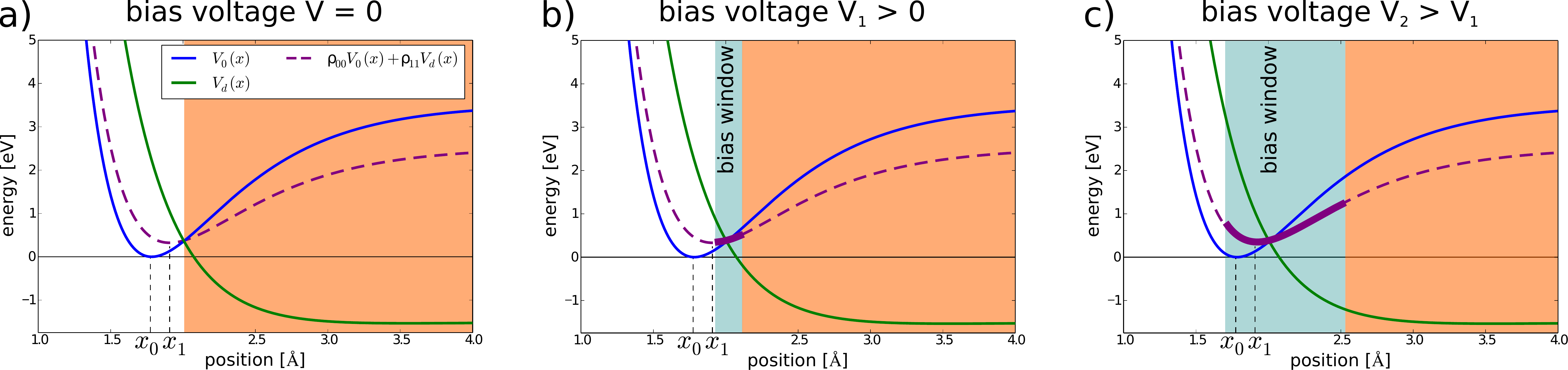}
		\caption{Potential energy surfaces and average potential $V_{\text{av}}(x)= \rho_{00} V_0(x)+ \rho_{11} V_d(x)$ for the model system stronger coupled to the right lead for three exemplary bias voltages $0 < V_1 < V_2$. The white area represents the {unpopulated regime}, the orange area is the {populated regime}, the blue shaded area highlights the {conducting regime}. The individual pictures a) to c) represent different applied bias voltages. As the bias is increased, the extent of the {conducting regime} enlarges and so is the zone of influence of the average potential.}
		\label{fig:asymm_explanation_voltage_window}
	\end{figure*}
	
	To explain the overall decreasing trend of the dissociation probability we first consider static nuclei and neglect partial electronic population by broadening effects. The {populated, unpopulated} and {conducting regime} as well as their dependence on the applied bias are identical to the symmetric coupling scenario considered in Sec.\ \ref{sec:symmetric}. In contrast to that, the average potential $V_{\text{av}}(x)= \rho_{00} V_0(x)+ \rho_{11} V_d(x)$ describing the nuclear motion for the molecule under resonant transport is modified by the changed molecule-lead coupling scenario. Notice that the minimum $x_1$ of the nonequilibrium potential $V_{\text{av}}(x)$ is now located in the {unpopulated regime} at zero bias (see Fig.\ \ref{fig:asymm_explanation_voltage_window} a). 
	As the bias is increased, the voltage window opens up in nuclear coordinate space, pushing the {populated regime} outwards. 
	As can be seen from Fig.\ \ref{fig:asymm_explanation_voltage_window}b -- c, the effect of the average potential $V_{\text{av}}(x)$ is to push the nuclei back to smaller $x$-values, thus counteracting dissociation. Therefore, the nucleus must reach larger $x$-values in order to dissociate, resulting in the overall decrease in dissociation probability upon increase of bias voltage as seen in Fig.\ \ref{fig:data_asymm__diss_Gamma} (middle). 
	In the adiabatic limit, this decrease of dissociation probability can be interpreted in terms of a built-up of a potential barrier around the interface of the {conducting} and the {unpopulated regime} with bias. 
	As such, a quantum-mechanical description of the nuclear degree of freedom may lead to corrections of the results obtained here within the Ehrenfest method. 
	This was considered by \citet{Dzhioev2011}, who have used the tunneling through voltage dependent adiabatic potential barriers to calculate the reaction rate for H$_2^+$ as a function of bias.

	An exception from the monotonous decrease of the dissociation probability is the result for $\Gamma=1.0$ eV, which  exhibits a local maximum around $V=2$ V. The local increase is characteristic for high molecule-lead coupling strengths as can be seen in Fig.\ \ref{fig:data_asymm__diss_Gamma} (bottom) for $1$ V -- $3$ V.
	The effect is caused by the broadening of the electronic level due to molecule-lead coupling and is therefore beyond the simplistic explanation based on different population regimes. Strong coupling $\Gamma$ leads to an enhanced partial population of the molecular electronic level and smears the border between the different charge regimes. These effects depend on the applied bias voltage and need to be compensated for by the force generated by the average potential along the extent of the {conducting regime}, leading to the  maximum in the dissociation probability at $1$ V -- $3$ V.

	The dissociation time (data not shown) is again on the order of $100$ fs and moderately increases with applied bias which is consistent with our interpretation of the change in the dissociation probability with bias.
	As before, the dissociation times are rather insensitive to $\Gamma$.
	
	Considering the average current through the system depicted in Fig.\ \ref{fig:data_asymm__diss_Gamma} (top), we find that the model allows for a pronounced resonant current above a certain bias voltage for any value of $\Gamma$. 
	This onset bias voltage lies in between the voltage at which the minimum of the conducting state $x_1$ enters the  {conducting regime} and the voltage at which the minimum of the unpopulated state $x_0$ leaves the 
	{unpopulated regime}. This observation is consistent with our interpretation of the behavior of the dissociation probability.
	
	Notice that for an asymmetric molecule lead coupling the results for dissociation probability and current will depend on the bias polarity. The results discussed above, obtained for positive bias voltage, are strongly influenced by the predominant coupling of the molecule to the right lead and the corresponding low population of the molecular electronic state. Upon reversing the bias polarity, the situation changes in such a way that the molecular electronic population is large. Consequently, the corresponding average potential gives rise to an optimal nuclear position which is located at large $x$-values, resulting in a dissociation probability (data not shown) that behaves more like the system in Sec.\ \ref{sec:symmetric} (or even like the system in Sec.\ \ref{sec:gate-voltage} if the coupling to the lead with the higher chemical potential becomes dominant).

\section{Conclusion}\label{sec:conclusion}

	We have investigated current-induced bond rupture in single-molecule junctions as a result of the transient population of anti-bonding electronic states by tunneling electrons.
	Applying a mixed-quantum classical approach to a generic model of a molecular junction, we have studied a wide range of physical parameters, ranging from the nonadiabatic regime of weak molecule-lead coupling to the adiabatic case of strong coupling as well as asymmetric coupling scenarios. We found that in certain parameter ranges a current across a molecular junction can not only induce the rupture of a chemical bond in the molecule, but under certain conditions it can also increase its stability.

	To rationalize these results we have introduced a concept, which employs a partitioning of the nuclear coordinate space in terms of the electronic population. 
	In order to understand the nuclear motion for a molecule under current, we considered the potential energy surface for a partially populated electronic level which is generated by the tunneling electrons.
	As long as the stable nuclear position for a system under current is located in the {populated regime}, an increase in bias voltage will push the nucleus outwards, thus increasing the probability for dissociation with bias.
	If this is not the case, however, an increase in bias voltage stabilizes the molecule. 
	The stability of molecules under current is an important aspect for possible future realizations of molecule-based nanoelectronic devices. 

\section*{Acknowledgement}
	We thank P.\ Auburger,  M.\ Bockstedte, and P.\ B.\ Coto for helpful discussions. This work was supported by the German Research Foundation (DFG) through SFB 953 and a research grant as well as the German-Israeli Foundation
	for Scientific Research and Development (GIF).

\bibliography{Bib}

\end{document}